\PassOptionsToPackage{table}{xcolor}
\documentclass[sigconf]{acmart}
\usepackage{cleveref}
\usepackage{ifthen}
\newboolean{revising}
\setboolean{revising}{false} 
\usepackage{subcaption}

\usepackage{xspace}
\usepackage{soul}

\PassOptionsToPackage{unicode}{hyperref}
\PassOptionsToPackage{naturalnames}{hyperref}
\hypersetup{
    colorlinks=true,
    linkcolor=blue,
    filecolor=black,
    citecolor=ForestGreen,
    urlcolor=ForestGreen,
}

\setboolean{revising}{false}
\ifthenelse{\boolean{revising}}
{
	
    \newcommand{\cut}[1]{\textcolor{gray}{\sout{#1}}}
\DeclareRobustCommand{\edit}[1]{\textcolor{blue}{#1}}
      
}{
	\newcommand{\edit}[1]{#1}
	\newcommand{\cut}[1]{}

    \newcommand{\tc}[1]{}
        
}

\newcommand{\ie}{\textit{i.e. }}
\newcommand{\eg}{\textit{e.g.,}}

\newif\ifcomments
\commentstrue   

\usepackage{booktabs}   
\usepackage{tabularx}
\newcolumntype{Y}{>{\raggedright\arraybackslash}X}

\newcommand{\arbor}{\textsc{Arboretum}\xspace}
\newcommand{\arbors}{\textsc{Arboretum}'s\xspace}

\newcommand{\DR}[1]{\textbf{DR#1}}

\usepackage{CJKutf8}

\usepackage{enumitem}
\usepackage{soul}
\usepackage{listings}
\usepackage{ulem}
\AtBeginDocument{%
  \providecommand\BibTeX{{%
    \normalfont B\kern-0.5em{\scshape i\kern-0.25em b}\kern-0.8em\TeX}}}

\copyrightyear{2026}
\acmYear{2026}
\setcopyright{cc}
\setcctype{by}
\acmConference[CHI '26]{Proceedings of the 2026 CHI Conference on Human Factors in Computing Systems}{April 13--17, 2026}{Barcelona, Spain}
\acmBooktitle{Proceedings of the 2026 CHI Conference on Human Factors in Computing Systems (CHI '26), April 13--17, 2026, Barcelona, Spain}
\acmPrice{}
\acmDOI{10.1145/3772318.3791656}
\acmISBN{979-8-4007-2278-3/2026/04}

\begin{document}

\title[Nonvisual Support for Understanding and Reasoning about Data Structures]{Nonvisual Support for Understanding and Reasoning about Data Structures}



\author{Brianna L. Wimer}
\email{bwimer@nd.edu}
\orcid{0000-0002-3821-5555}
\affiliation{
    \institution{University of Notre Dame}
    \city{Notre Dame}
    \state{Indiana}
    \country{USA}
}

\authornote{The first two authors are equal contributors to this work.}

\author{Ritesh Kanchi}
\email{rkanchi@g.harvard.edu}
\orcid{0009-0006-7978-0821}
\affiliation{
    \institution{Harvard University}
    \city{Cambridge}
    \state{Massachusetts}
    \country{USA}
}
\authornotemark[1]

\author{Kaija Frierson}
\email{kaijaf@uark.edu}
\orcid{0009-0005-0961-7618}
\affiliation{
    \institution{University of Arkansas}
    \city{Fayetteville}
    \state{Arkansas}
    \country{USA}
}

\author{Venkatesh Potluri}
\email{potluriv@umich.edu}
\orcid{0000-0002-5027-8831}
\affiliation{
    \institution{University of Michigan}
    \city{Ann Arbor}
    \state{Michigan}
    \country{USA}
}

\author{Ronald Metoyer}
\email{rmetoyer@nd.edu}
\orcid{0000-0003-2206-1720}
\affiliation{
    \institution{University of Notre Dame}
    \city{Notre Dame}
    \state{Indiana}
    \country{USA}
}

\author{Jennifer Mankoff}
\email{jmankoff@cs.washington.edu}
\orcid{0000-0001-9235-5324}
\affiliation{
    \institution{University of Washington}
    \city{Seattle}
    \state{Washington}
    \country{USA}
}

\author{Miya Natsuhara}
\email{mnats@cs.washington.edu}
\orcid{0009-0008-7464-3670}
\affiliation{
    \institution{University of Washington}
    \city{Seattle}
    \state{Washington}
    \country{USA}
}

\author{Matt X. Wang}
\email{mxw@cs.washington.edu}
\orcid{0009-0003-0708-9378}
\affiliation{
    \institution{University of Washington}
    \city{Seattle}
    \state{Washington}
    \country{USA}
}

\renewcommand{\shortauthors}{Wimer and Kanchi et al.}



\begin{abstract}
Blind and visually impaired (BVI) computer science students face systematic barriers when learning data structures: current accessibility approaches typically translate diagrams into alternative text, focusing on visual appearance rather than preserving the underlying \cut{abstractions}\edit{structure} essential for \cut{algorithmic reasoning}\edit{conceptual understanding}. More accessible alternatives often do not scale in complexity, cost to produce, or both. Motivated by a recent shift to tools for creating visual diagrams from code, we propose a solution that automatically creates accessible representations from structural information about diagrams. \edit{Based on a Wizard-of-Oz study, we derive design requirements for an automated system, }\cut{We articulate four principles for accessible data structure representations to preserve these properties: semantic roles, navigational affordances, structural reasoning, and multimodality.}\cut{Our system,}\arbor, \cut{operationalizes these principles by generating synchronized accessible representations}\edit{that compiles text-based diagram specifications into three synchronized nonvisual formats}—tabular, navigable, and tactile\cut{—from diagram specifications}. Our evaluation with BVI \cut{learners}\edit{users} highlights the strength of tactile graphics for complex tasks such as binary search; the benefits of offering multiple, complementary \edit{nonvisual} representations\cut{ rather than a single modality}; and limitations of existing digital navigation patterns for structural reasoning. This work reframes \cut{accessible representations of}\edit{access to} data structures \cut{around}\edit{by} preserving \edit{their structural}\cut{computational} properties. \edit{The solution is} \cut{providing both design principles and} a practical system to advance accessible CS education.

\end{abstract}

\ccsdesc[500]{Human-centered computing~Accessibility}
\ccsdesc[500]{Social and professional topics~Computing education}

\keywords{Accessibility, Diagramming, Computer Science Education}

\begin{teaserfigure}
\vspace{1em}
 \centering
  \includegraphics[width=0.8\textwidth]{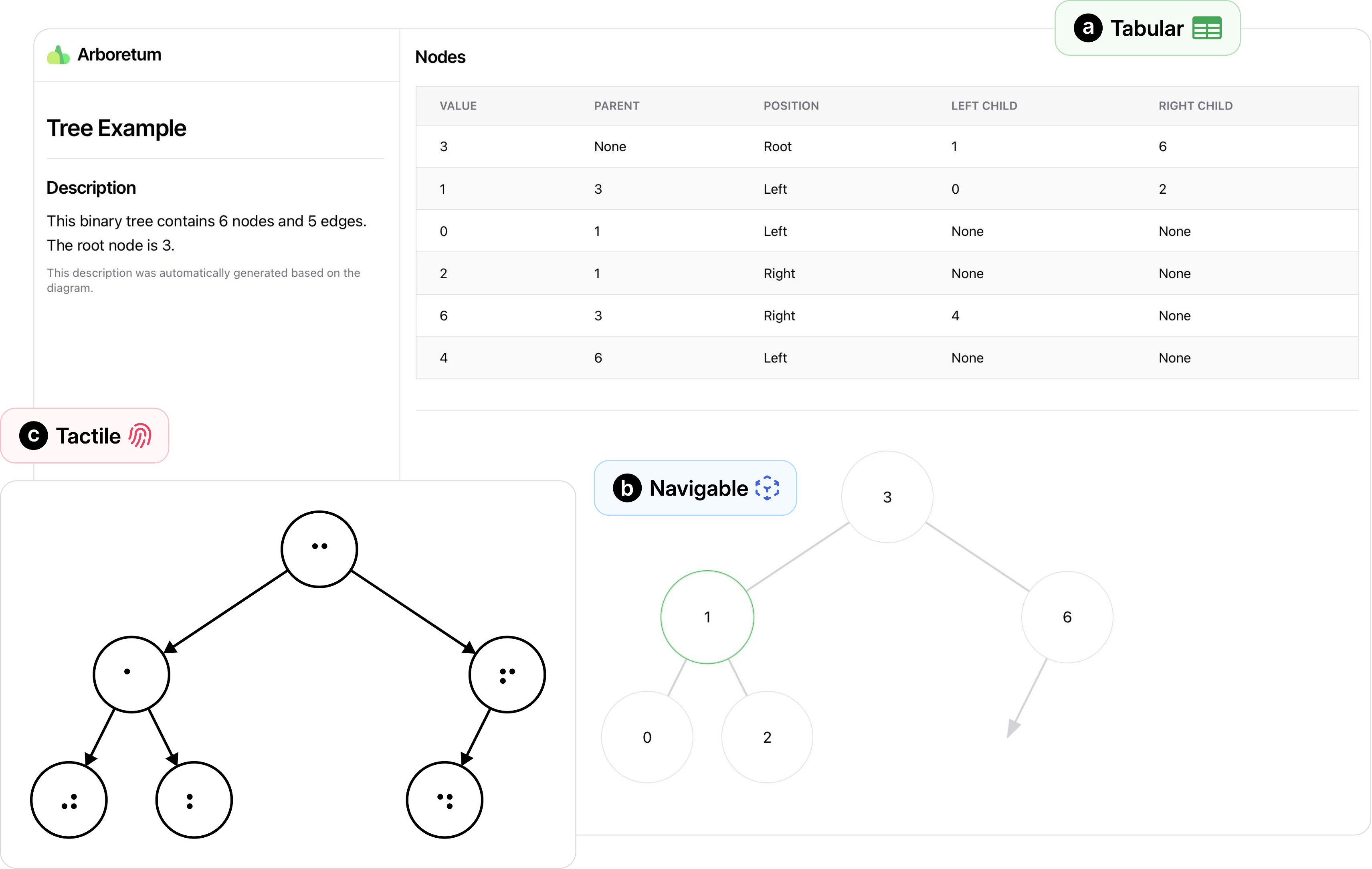}
  \caption{\arbor, a web-based system that generates accessible diagrams of introductory data structures in (a) tabular, (b) navigable, and (c) tactile representations. This example highlights a binary tree in \arbor across all three representations.}
  \Description{A screenshot of Arboretum's Preview mode with a binary tree example. The example shows a binary tree with six nodes, five edges, and a root of 3. The three accessible representations are also shown: (a) a tabular representation, with a table listing the binary tree's nodes; (b) a navigable representation, with a screen reader-navigable tree; and (c) a tactile representation, with circular Braille-labeled nodes connected by arrowed edges.}
  \label{fig:teaser}
  \vspace{1em}
\end{teaserfigure}


\maketitle

\begin{CJK*}{UTF8}{gbsn}
\section{Introduction}
The road into computer science (CS) is paved with visual representations \cite{fouh2012role}.\cut{ From the classroom to professional practice, visual representations are}\edit{ They are} central to reasoning about computing: programmers sketch designs to reason about their software and how their data is structured  \cite{hayatpur23cross,cher07whiteboard,yatani09diagram}, and CS instructors rely on diagrams, whiteboard sketches, and animations to communicate the complexity of core concepts \cite{fouh2012role,fincher20notional}. In introductory CS education \edit{(\eg{} CS1, CS2),}\cut{is} data structures such as arrays and binary trees are often introduced through \edit{such} diagrams,\cut{that} aid\edit{ing} abstraction, support\edit{ing} mental model development, and reduc\edit{ing} cognitive load \cite{tippett2016recent,hayatpur23cross,torres20191approaches}. However, these representations remain inaccessible to blind and visually impaired (BVI) students \cite{torres20191approaches}, who already face barriers in CS education due to under-prepared educators and the limited availability of accessible instructional materials \cite{ladner2015increasing, ladner2016all}. 

\begin{figure}
    \begin{subfigure}[t]{0.3\textwidth}
    \vspace{2em} 
    \includegraphics[width=0.99\linewidth]{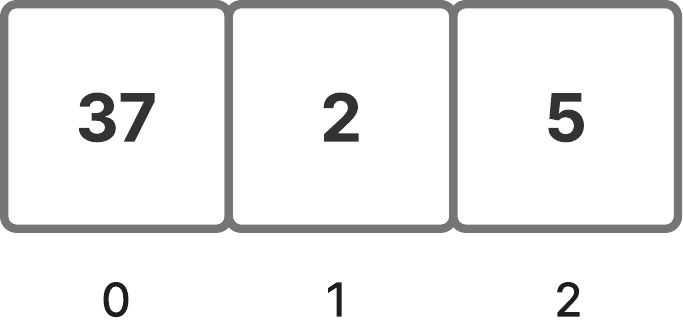}
    \Description{A diagram of an array in a slide of a slide deck, comprised of three adjacent blocks containing corresponding values, with the index beneath each.}
    \caption{Array.}
    \label{fig:og-array}
    \end{subfigure}    
    \begin{subfigure}[t]{0.4\textwidth}
     \vspace{2em} 
    \includegraphics[width=0.99\linewidth]{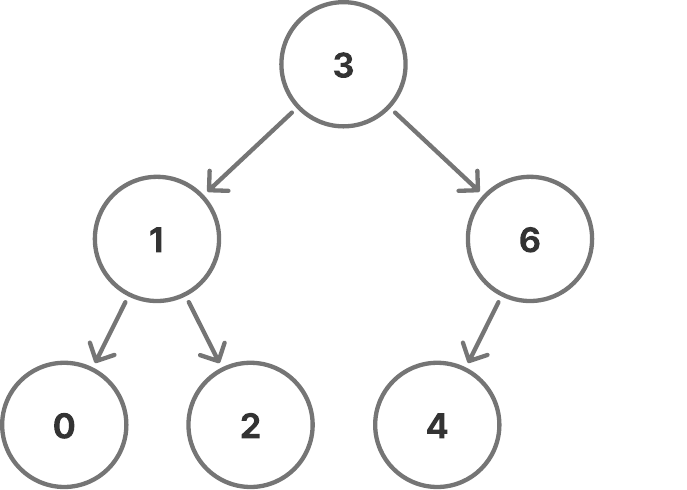}
    \Description{A diagram of a binary tree in a slide of a slide deck, comprised of nodes connected by arrowed-edges.}
    \caption{Binary Tree.}
    \label{fig:og-tree}
    \end{subfigure}
    \Description{Visual diagrams of an array and binary tree. The array is comprised of three adjacent blocks containing corresponding values, with the index beneath each. The binary tree is comprised of nodes connected by arrowed-edges.}
    \caption{Representations of standard array and binary tree diagrams as seen in a slide deck.}
    \label{fig:og-viz}
\end{figure}

\edit{To make the problems BVI learners face more concrete, consider a hypothetical student, Maya, exploring an array presented visually as a row of adjacent boxes, each containing a value and labeled with its index (Figure~\ref{fig:og-array}). A hurried instructor, doing their best to follow Web Content Accessibility Guidelines (WCAG) for ``complex images'' (\eg{} charts, graphs, diagrams, illustrations, maps) \cite{wcag21}, provides an alternative text description (\eg{} ALT text) that simply lists the values: ``5, 2, 9, 7\ldots''. To identify the element at array index 6, Maya must count sequentially through each value, unlike her sighted peers who can can immediately see which value sits in the box labeled with index ``6.'' This strategy quickly becomes cognitively intensive for BVI students as arrays grow in size, and instructors often struggle to maintain consistent nonvisual descriptions as diagrams become larger or get updated shortly before class. A similar issue arises with node-link diagrams for binary trees: sighted learners rely on spatial cues---circles connected by arrows with parents above children and left/right positioning (Figure~\ref{fig:og-tree})---while Maya receives ALT text that states: ``The binary tree has a root with the value 4. The 4's children have values of 2 and 8. The 2's child has a value of 1. The 8's children have values of 6 and 10.'' Even this level of detail varies depending on the instructor’s expertise or time to prepare.}
\edit{For Maya, reasoning about operations like binary search or verifying the binary search tree (BST) property results in an over-reliance on working memory or repeated passes over the description to decipher semantic roles and relationships. ALT text often omits\cut{what we call} the \textit{computational properties} of data structures: the rules, relationships, and reasoning procedures that define how a structure is organized and how it can be traversed or manipulated.}
\edit{Although Maya’s experience reflects the minimum legal standard for classroom accessibility \cite{DOJ}, it falls short of providing equitable access: diagrams are treated as images to be described rather than as structured representations that can be encoded. \citet{larkin1987diagram} note that diagrams support reasoning by organizing information spatially, yet this structure is lost when diagrams are exported as images and later paired with text. Supporting equitable access therefore requires representing diagrams in a form where structural relationships are captured directly rather than left implicit.}

Prior research has solved this problem by developing alternatives to text. \edit{Tactile representations can preserve layout and relationships,} \cut{data structure elements and support}\edit{while supporting} conceptual understanding of CS concepts for BVI learners \cite{improving-understanding, alzalabny2024touch, polzer2014tactile}. However, producing tactile materials requires time and expertise and is manually done separately for each new diagram, \edit{making them challenging to create consistently for instruction} \cite{defining-problems-tactile, exploring-perspectives-teachers,improving-understanding,prescher2014production}. 
\edit{If structural information about diagrams, instead of images of diagrams, were available, it would be feasible to automate this process, as well as automating ALT text generation and other accessibility solutions. An automated solution could vastly improve the speed and scale at which diagrams could be made accessible.} 
\edit{Structural information is often available in some form during the diagram creation process. Some instructors use tools to create diagrams, ranging from general-purpose presentation software (e.g., PowerPoint, Google Slides) to specialized diagramming environments (e.g., draw.io, Lucidchart). However, each tool introduces its own stylistic conventions and offers limited accessibility features.}
\edit{Others use text-based diagram specifications (\eg{} Mermaid\footnote{\url{https://mermaid.js.org}}, Graphviz DOT\footnote{\url{https://graphviz.org/doc/info/lang.html}} PlantUML\footnote{\url{https://plantuml.com/}}, d2\footnote{\url{https://d2lang.com/}}).} 
\edit{Our key insight is that \textbf{these tools naturally capture structural information including nodes, relationships, and reasoning patterns in a single source of truth from which accessible multimodal representations can be automatically generated.}}

\edit{Building on this insight, we present the iterative design of an automated solution for making diagrams accessible.} \cut{in this work, We focus}\edit{Our work focuses} on arrays and binary trees,\cut{both of which are} foundational \edit{data structures} in introductory CS\cut{education}: arrays demonstrate sequential ordering and index-based access, while binary trees demonstrate different traversal strategies \cite{data-structures-and-their-algos}. 
\edit{We have three primary contributions. First, we \textbf{derive five design requirements  from a small Wizard-of-Oz study of three nonvisual prototypes of general-purpose flowcharts}: textual with Q\&A support, navigable, and tactile formats. Our goal was to understand accessibility barriers across representation modalities. Participants had difficulty forming complete mental models from any single prototype, and different prototypes had complementary strengths and weaknesses.}
\edit{The resulting requirements shaped the design and implementation of \arbor,} \textbf{a web-based system that compiles diagram specifications into tabular, navigable, and tactile representations that make a data structure's computational properties explicit} (Figure~\ref{fig:teaser}). \cut{These representations provide accessible alternatives to visual diagrams by preserving the structural layout and relationships among elements \cite{alzalabny2024touch}, while enabling learners to explore structures through touch actively, reinforcing conceptual understanding \cite{improving-understanding}.} \edit{\arbor automates this process through access of structured data rather than images.}
\edit{To validate \arbor, we} asked 8 BVI learners to use \arbors representations. Participants successfully worked through five task types: (1) locating specific elements, (2) identifying parent–child relationships, (3) identifying leaf nodes, (4) verifying BST properties, and (5) performing binary search.  We show that the \textbf{nonvisual representations generated by \arbor  support comprehension and reasoning over arrays and binary trees.} Based on these contributions, we highlight opportunities for expanding the concepts to a wider array of educational settings.

\section{Related Work}
Our research \cut{sits at the intersection of}\edit{intersects} two strands of prior work: accessible CS education for BVI learners, which highlights \cut{both}the barriers students face and the systems developed to support them; and accessible diagrams in STEM, which examines tactile, auditory, and multimodal alternatives to visual representations.

\subsection{Accessible CS Education and Systems for BVI Learners}
BVI learners encounter a range of visually-oriented barriers in CS education---spanning curriculum, tools, and classroom practices \cite{educational-exp-blind}. Current computing education environments often lack strategies for making learning both inclusive and accessible, though educators and researchers have explored approaches to address these gaps.

At the curriculum level, efforts include the creation of screen reader-friendly instructional content and physical components that embody abstract computing concepts \edit{\cite{tac-teaching-inclusively, teachaccess, accesscomputing, kanchi2025systems}}. Visual representations---commonly used to teach data structures, algorithms, and other programming concepts---continue to present significant barriers for BVI learners when not paired with accessible alternatives \cut{CS}\cite{luque2017inclusion, reducing-blind, educational-exp-blind, addressing-access-visual-impair}. To counter this, schools for the blind and visually impaired have long used tactile manipulatives--boxes to represent variables, dice for numerical values, and switches for Boolean values \cite{design-bvi-cs}. Additional work has focused on curriculum redesign efforts: for example, \citet{computer-science-principles-teachers-bvi} adapted the Code.org AP Computer Science Principles course by offering accessible and unplugged activity alternatives and ensured that content adhered to WCAG 2.1 AA \cite{wcag21} and ARIA \cite{waiaria} guidelines.

Some systems have focused on making computing concepts accessible through alternative representations. For example, the \textit{GSK} system \cite{balik2014gsk,balik2013gsk} enables BVI and sighted learners to create, edit, and share graphs using familiar interaction mechanisms (\eg{} mouse, keyboard, screen reader). To support both groups, GSK provides synchronized grid and connection views of a graph, ensuring proper representation and interaction. 

Beyond CS concepts, other systems have aimed to make \cut{computing and}programming environments\cut{more} accessible. For instance, \citet{javaspeak-programming-tool} introduced \textit{JavaSpeak}, a specialized Java programming environment with aural feedback that ``speaks'' a program's semantics and structure analogous to how syntax highlighting and indentation visually communicate structure to a sighted learner. Similarly, \textit{Sodbeans} offers a computer programming environment with custom screen reader, \textit{Hop} programming language, and accompanying talking debugger \edit{\cite{design-bvi-cs, stefik2011empiricalstudiesonpl}.} \edit{Other accessible programming languages include Blocks4All \cite{accessible-to-whom} and Quorum \cite{stefik2017quorum}. At a broader level of abstraction, \citet{Schanzer:2019:AccessibleAST} developed a language-independent toolkit for creating accessible programming environments, providing spoken descriptions and accessible navigation through a unified block editor for BVI programmers. Plugins to adapt existing IDEs have also been developed; \textit{StructJumper,} an Eclipse plugin that exposes the \edit{structure} of a Java class via a hierarchical tree for easy code navigation \cite{structjumper} and \textit{CodeTalk}, a Visual Studio plugin which addresses accessibility issues in code comprehension, editing, debugging, and collaboration for the BVI developer experience \cite{potluri2018codetalk}.}

Collectively, these curricular efforts and systems demonstrate that accessible design and alternative representations are critical for BVI learners in CS. They also highlight that accessibility must be built into both instructional materials and educational tools. Our system, \arbor, builds on this foundation by focusing on data structure diagrams, providing a low-barrier interface for educators and students to create and explore accessible representations.



\subsection{Nonvisual Presentations of Data from Diagrams}
Beyond CS education settings, researchers have explored various nonvisual approaches, from tactile graphics to multimodal audio-tactile systems, to preserve the essential properties of diagrams across computing contexts including flowcharts \cite{alzalabny2024touch,cross2020transforming, reducing-blind}, node–link diagrams \cite{zhao2024tada,kennel96audiograf,fan2022accessibility}, UML diagrams \cite{teaching-uml,luque2014can,wildhaber2020uml,doherty2015uml,petrie2002tedub,petrie2006providing,king2004presenting,horstmann2004automated}, and graphs \cite{balik2014gsk,syal2016digvis,cohen06plumb,calder06plumb,cohen2006teaching,balik2013gsk}.

Tactile representations preserve spatial layouts and relationships through physical formats such as raised-line graphics on swell paper thermoformed materials, as well as digital refreshable tactile displays \cite{Gardner1996TactileGA,polzer2014tactile,prescher2014production}. For \cut{example}\edit{instance}, hyperbraille displays have enabled users to directly follow interconnections in mind maps via raised nodes and lines \cite{polzer2014tactile}. While tactile representations excel at conveying spatial structure and supporting direct manipulation, they face significant scalability challenges: physical production requires specialized equipment and expertise \cite{prescher2014production}.

Researchers have attempted to preserve a visual representation's structure through auditory approaches that map to screen reader navigation patterns \edit{\cite{polzer2013making,blanco2022olli,elavsky-data-nav,zhao2004sonification,thesonificationhandbook}}, but these sacrifice the spatial grounding that makes tactile representations effective. Multimodal representations emerged to combine the spatial grounding of tactile exploration with detailed information delivered auditorily, as users often value this parallel presentation of information across modalities \cite{reducing-blind,wildhaber2020uml,topsen-comparing-code-strats}. Early work by Blenkhorn and Evans \cite{blenkhorn1998using} created audio-tactile matrices for data flow diagrams, demonstrating how combined modalities could support both spatial layout and content detail. Building on this foundation, researchers have developed various interaction approaches: Kennel et al. \cite{kennel96audiograf} enabled users to trigger sounds through touch, Alzalabny et al. \cite{alzalabny2024touch} augmented two-dimensional tactile interfaces with audio feedback for technical diagrams, and Kawulok et al. \cite{reducing-blind} implemented gesture-driven audio descriptions. More recent tablet-based systems allow users to navigate through touch while receiving auditory cues to identify objects and interconnections \cite{zhao2024tada,wildhaber2020uml}. These systems all enable ease for constructing diagrams, taking advantage of multimodality for nonvisual presentations of data.



Evaluating nonvisual representations has become increasingly important as approaches have diversified. Since nonvisual representations vary widely and offer different benefits, prior work has examined how to compare their efficiency to access information \cite{watanabe2018effectiveness,watanabe2012development,yu2002multi}, user preferences \cite{goncu2010usability}, representation usability \cite{yang20tactile,goncu2010usability}, workload \cite{yu2002multi}, and user performance across tasks \cite{yang20tactile,goncu2010usability,yu2002multi,engel2018user}. While prior research focuses on general-purpose diagrams, there is little systematic understanding of how nonvisual representations support learning and conceptualizing domain-specific diagrams in computing remains limited. Our work addresses this by focusing specifically on data structure diagrams and operationalizing access through complementary representations that preserve computational meaning.

\section{Wizard-of-Oz Study for Representation Design}
\label{sec:feasibility}
\edit{The goal of our first study was to understand the tradeoffs between different representations of a generic class of frequently used diagrams: node-link diagrams (specifically, flowcharts). These diagrams use the same node-link structure as data structure diagrams, which appear widely in CS education \cite{fouh2012role}. Using a Wizard-of-Oz approach allowed us to explore tradeoffs between representations before fully implementing them \cite{dahlback1993wizard}.}

\paragraph{Representations}
\edit{We created three nonvisual prototypes: 1) an ALT-text description with a high-level overview and a linearized list of nodes and edges, supplemented by an LLM-based question-answering component; 2) a keyboard-navigable digital graph whose custom navigation followed the diagram’s directional flow (necessary because flowcharts with cycles do not map cleanly onto existing screen reader patterns); and 3) a tactile diagram printed on swell paper with a digital legend. We implemented a back-end system supporting our Wizard-of-Oz study of the Q\&A interface and the digital graph. When participants wanted to ask questions, or explore the digital graph, they verbalized their intent and researchers executed the commands needed to generate spoken output. Additionally, since four of our participants were unfamiliar with reading braille, researchers provided verbal label information on request for the tactile diagram. }


\paragraph{Method and Analysis}
\edit{We described the 
study's purpose and procedures, and introduced flowchart-related
concepts and terminology to participants. We then asked them to explore the three prototypes,
starting with the ALT text representation. After exploring each prototype, participants
answered usability questions and provided feedback on their experience. Sessions were about 90-minutes long and took place in-person at the \textbf{(anonymized location and organization for review)}. All sessions were recorded with consent, and participants received a \$60 Amazon
gift card as compensation. 
Recordings and transcripts were analyzed using an affinity diagramming process \cite{scupin1997kj}. We focused on the usability of each prototype, their effectiveness in assisting participants in answering
flowchart-related questions, and identifying areas for enhancement based on participant
feedback.}

\paragraph{Participants}
\edit{
Eight participants, several who had prior experience with flowcharts, were recruited through a national organization supporting BVI professionals.  
Ages ranged from 36 to 67 (mean = 51 years, SD = 12). We refer to these participants as FP1-FP8 throughout our findings.}




\subsection{Findings and Resulting Design Requirements}
\label{sec:rq}
\edit{This first study revealed five recurring accessibility needs for diagram representations:}

\paragraph{DR1. Adopt standardized screen reader navigation models.}
\edit{
We implemented a Wizard-of-Oz navigation pattern that allowed participants to move directionally along the diagram’s logical flow (\eg{} advancing through outgoing edges, returning via incoming edges, and following labeled branches) rather than relying on a parent–child hierarchy. Although this pattern matched the structure of the flowcharts, participants had difficulty using it; when sibling nodes reflected different decision paths or cycles brought them back to earlier nodes, the unfamiliar commands became hard to interpret. In contrast, participants readily used ARIA-style commands when they applied, drawing on existing expertise. As FP2 noted: ``Using the keystrokes is easy\ldots knowing that parent was left, child was up and down\ldots that makes it easier.'' Such reactions guided us toward standardizing navigation to match familiar screen reader interaction models, reducing both the learning burden and interaction friction.}
\paragraph{DR2. Provide structural information in tables.}
\edit{ Participants consistently suggested presenting nodes and edges in a tabular structure instead of a list inside prose because it aligns with existing screen reader expertise in table navigation: ``Using arrow keys with screen readers and spreadsheets\ldots it's much easier for me. I can quickly go back and forth'' (FP8). Tables offer a stable, predictable baseline format for making data accessible \cite{zong2022rich, wang2024table}---a format that separates structural facts from explanation, supports random access, and reduces the variability inherent in authored descriptions.}

 \paragraph{DR3. Explicitly encode structural relationships rather than requiring inference.} 
\edit{Participants had difficulty determining how elements were connected when relationships were described in the ALT text format; they lacked clear cues about which nodes were connected or how they were positioned relative to one another. Because the ALT text listed nodes and edges linearly, participants could not determine which elements were related without reconstructing structure through inference. One participant stated: `Trying to understand which nodes are attached to the edges
and the different labels was somewhat difficult.'' (FP3), while another reported: ``I wasn't really getting the layout or structure'' (FP8).}



\paragraph{DR4. Separate structural facts from narrative explanation in nonvisual representations.} 
\edit{Participants struggled with the ALT text prototype because moving back and forth between a high level overview and a lists of nodes and edges made it difficult to track what information was explanatory and what was structural. As one participant stated, ``You get this big data dump…trying to break it down is hard'' (FP7). Others reported difficulty understanding the ordering of details, with FP1 noting, ``I wasn’t getting a cohesive picture…the order of things wasn’t clear.'' Representations should therefore separate conceptual framing from structural detail, providing a clear overview alongside a distinct, precisely structured enumeration of nodes, edges, and roles.} 


\paragraph{DR5. Enable integrated access across complementary modalities.}

\edit{No single representation contained enough information for participants to understand the full structure; when restricted to one representation, they often felt uncertain or confused. Instead, participants preferred when they combined representations, using each to compensate for the limitations of another. Participants explicitly described integration strategies: ``I'd use a tactile printout to verify the visual map in my head'' (FP8) and ``I would check [the Q\&A answers] by verifying with the raw data table'' (FP1). Here, ``integrated access'' refers to supporting smooth movement between representations and enabling users to cross-check and coordinate information across modalities.
}

\edit{Together, these requirements expose the structural information that is normally implicit in visual diagrams and motivate synchronizing multiple modalities and representations. 
}

\section{The \arbor System}
\label{sec:arbor}

\cut{\mbox{\arbor} is a web-tool designed to address lossy translation in accessible data structure representations within introductory CS environments. }


\edit{\arbor is a web-based system that generates accessible data structure---currently array and binary tree---diagrams for introductory CS environments, ensuring that structural information is preserved\footnote{A live demo can be accessed at \href{https://g.riteshkanchi.com/chi26-arboretum/demo}{g.riteshkanchi.com/chi26-arboretum/demo}}.}\cut{\arbor Rather than converting visual diagrams into a single alternative format, \arbor operationalizes our guiding principles across three synchronized and multimodal representations---tabular, navigable, and tactile.} \edit{\arbor operationalizes the design requirements elicited in our Wizard-of-Oz study by compiling diagrams written in diagram specification languages into three synchronized representations: tabular, navigable, and tactile.} By ``synchronized,'' we mean that all representations are generated from the same underlying structural model, ensuring they remain consistent and convey the same structural information. \edit{Rather than implementing many-to-many translators, we compile the text-based diagram specifications into an intermediate representation (IR) to provide a single unified format and infer additional accessibility properties of the data structure.}


\arbor was developed with both CS educators and students in mind: \textbf{educators} who need to produce accessible data structure diagrams for instruction, and \textbf{students}, who\cut{use the accessible outputs to learn and navigate data structures} \edit{engage with multimodal outputs to explore and learn data structures accessibly.} \edit{There are many text-based diagram specification languages which describe nodes, edges, and layout rules in a simple textual syntax that a renderer converts into a visual diagram. Typically, these languages offer limited accessibility support, often only allowing authors to add manual ALT text.} By directly supporting \edit{familiar diagram specification languages}, \arbor makes accessibility effectively zero-cost for pre-existing diagrams\cut{ in these languages}, requiring minimal additional work from diagram creators.

\cut{In this work, we focus on two fundamental data structures: arrays and binary trees \cite{data-structures-and-their-algos}. These structures allowed us to demonstrate \arbors features and evaluate our \cut{principle-guided} output representations for BVI learners. }


\subsection{Implementation}\label{sec:arbor-implementation}

\begin{figure*}
    \centering
    \includegraphics[width=0.99\linewidth]{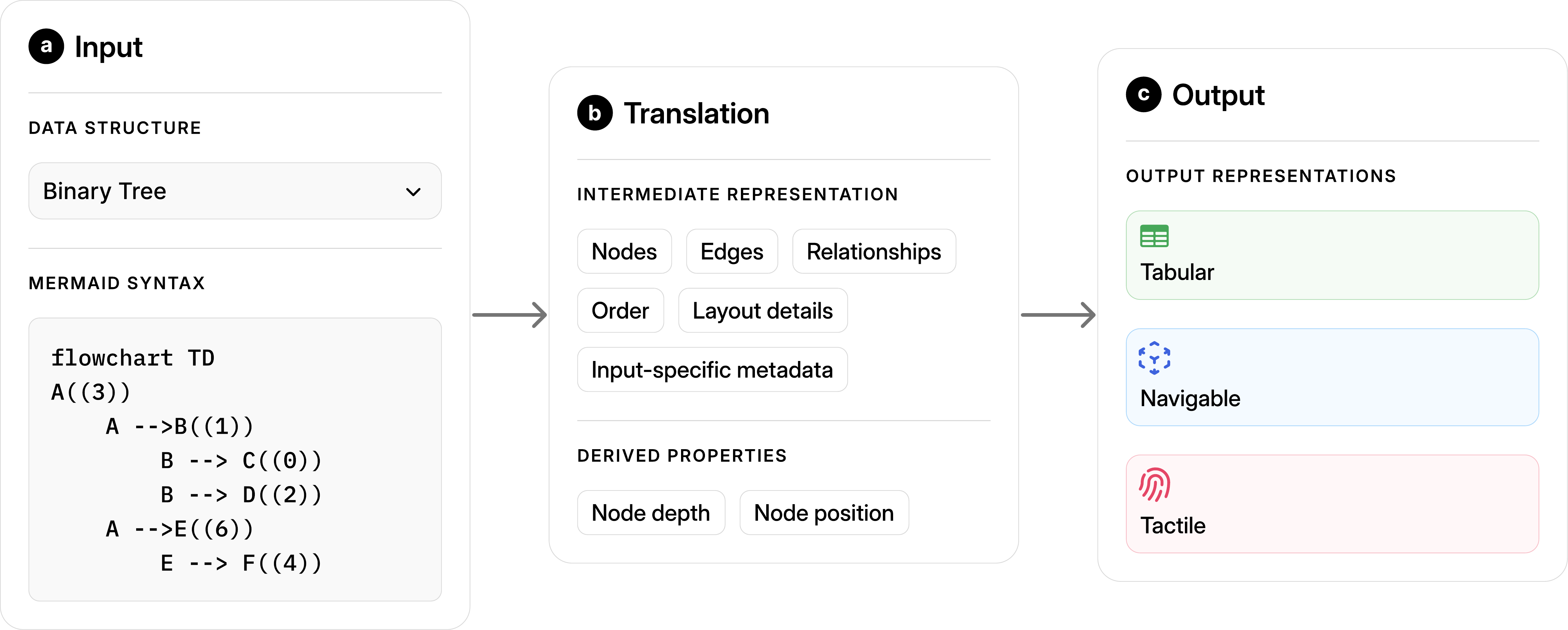}
    \Description{This figure illustrates \arbors three-stage pipeline from left to right.
    Input (Stage (a)) shows "Binary Tree" as the selected data structure and the following Mermaid syntax:
    flowchart TD; A((3)); A -->B((1)); B --> C((0)); B --> D((2)); A -->E((6)); E --> F((4)).
    Translation (Stage (b)) parses, extracts, and compiles the received syntax into an IR with nodes, edges, relationships, order, layout details, and input-specific metadata. It also includes derived properties such as node depth and node position.
    Output (Stage (c)) shows three output representations modalities: Tabular, Navigable, and Tactile.
    }
    \caption{A system overview of \arbor: an (a) input stage with a selected data structure and a provided Mermaid diagram specification, a (b) translation stage compiles the input into the IR, extracting and deriving additional properties, and an (c) output stage that generates accessible output representations.}
    \label{fig:system-diagram}
\end{figure*}

\arbor is a client-side web application developed with TypeScript, React, and Next.js\footnote{\url{https://nextjs.org}}\footnote{The source code is available at \href{https://github.com/ritesh-kanchi/Arboretum}{github.com/ritesh-kanchi/Arboretum}}. \edit{The system}\cut{\mbox{\arbor}} follows a three-stage pipeline: \edit{input specification, translation, and output generation} (Figure~\ref{fig:system-diagram}). \edit{\arbor introduces an IR that encodes the accessibility-relevant semantics (\eg{} traversal order, role labeling, explicit structural relationships) needed for accessible data structure representations. Input diagram specifications are compiled into this IR, and a separate compiler pass generates the accessible outputs from it.} 


\paragraph{Input} Some instructors visually author diagrams in direct manipulation tools (\eg{} PowerPoint, Lucidchart, Canva), while others use text-based diagram specification languages to use declarative syntax that supports abstraction and reuse in ways that direct manipulation tools do not \cite{pollock2024bluefish}. \arbor accepts diagrams from some of the more popular of these text-based languages, specifically Mermaid \edit{(Figure~\ref{fig:ir-tree-example-mermaid})} and Graphviz DOT \edit{(Figure~\ref{fig:ir-tree-example-graphviz}).} \cut{for data structures, enabling educators to keep their existing workflows while automatically generating accessible materials.}\cut{---currently arrays \cut{or binary trees---} We focus on these two diagram specification languages because they are commonly used in STEM education and offer clean, declarative syntax.} \cut{The system’s translation stage requires well-defined structure, so}\edit{\arbor also requires that} authors must specify the input's data structure type to ensure \edit{proper compilation into the IR (Figure~\ref{fig:system-diagram}a).} \cut{the translation stage compiles properly}.

\begin{figure}[hbt!]
  \centering
  \begin{subfigure}[t]{0.2\textwidth}
    \vspace{0pt} 
    \begin{verbatim}
flowchart TD
A((1))
    A -->B((2))
        B --> C((3))
        B --> D((4))
    A -->E((5))
        E --> F((6))
    \end{verbatim}
    \caption{Mermaid syntax.}
    \label{fig:ir-tree-example-mermaid}
  \end{subfigure}
  \hfill
  \begin{subfigure}[t]{0.25\textwidth}
    \vspace{0pt}
    \begin{verbatim}
digraph bt {
    A[label="1"];
    B[label="2"];
    C[label="3"];
    D[label="4"];
    E[label="5"];
    F[label="6"];

    A->B;
        B->C;
        B->D;
    A->E;
        E->F;
}
    \end{verbatim}
    \caption{Graphviz DOT syntax.}
    \label{fig:ir-tree-example-graphviz}
  \end{subfigure}
  \hfill
  \begin{subfigure}[t]{0.45\textwidth}
    \vspace{0pt}
    \begin{verbatim}
ArborBinaryTree:
  meta:
    type = "binary_tree"
    ...    # optional metadata
  
  nodes = [
    Node(id = "A", value = "1", ...)
    Node(id = "B", value = "2", ...)
    ...    # additional nodes
  ]

  edges = [
    Edge(parent = "A", child = "B", ...)
    Edge(parent = "B", child = "C", ...)
    ...    # additional edges
  ]
    \end{verbatim}
    \caption{Intermediate representation.}
    \label{fig:ir-tree-example-ir}
  \end{subfigure}
  \Description{This figure details both Mermaid (a) and Graphviz DOT (b) input syntax for the same binary tree, which is then compiled into Arboretum’s shared intermediate representation (c). The syntaxes and intermediate representation are in columns from left to right.}
  \caption{\edit{Examples of Mermaid (a) and Graphviz DOT (b) input syntax for the same binary tree compiled into \arbors shared IR (c), which unifies distinct diagram formats into a shared, semantically consistent model.}}
  \label{fig:ir-tree-example}
\end{figure}

\paragraph{Translation} In the translation stage, \arbor \cut{parses the DSL into an abstract syntax tree with the key components of the diagram---its elements and how they connect  (Figure~\ref{fig:system-diagram}b).} \edit{compiles inputs into the IR by first parsing the source specification into an abstract syntax tree that captures its elements and their connections (Figure~\ref{fig:system-diagram}b). Because diagram specification languages can differ in syntax and expressive power, each compiler pass is tailored to the source language (\eg{} Mermaid or Graphviz DOT) and the data structure type the author specifies.} 
\cut{Because diagram specifications vary in detail and precision, this stage also performs semantic augmentation, inferring missing roles, relationships, or levels to ensure completeness.}  
\edit{Figure~\ref{fig:ir-tree-example} shows how Mermaid and Graphviz specifications of the same binary tree are compiled to the same IR. Figure~\ref{fig:ir-example} illustrates how the IR supports arrays and naturally generalizes to linked lists and two-dimensional arrays.}

\begin{figure}[hbt!]
  \centering
\begin{subfigure}[t]{0.45\textwidth}
    \vspace{0pt}
    \begin{verbatim}
ArborArray:
  meta:
    type = "array"
    ...    # optional metadata
  
  elements = [
    Element(id = "A", value = "1", ...)
    Element(id = "B", value = "2", ...)
    ...    # additional elements
  ]
    \end{verbatim}
    \vspace{-0.5em}
    \caption{Array.}
    \label{fig:ir-example-arr}
  \end{subfigure}
  \hfill
\begin{subfigure}[t]{0.45\textwidth}
    \vspace{0pt}
    \begin{verbatim}
ArborLinkedList:
  meta:
    type = "linked_list"
    ...    # optional metadata
  
  nodes = [
    Node(id = "A", value = "1", ...)
    Node(id = "B", value = "2", ...)
    ...    # additional nodes
  ]
    \end{verbatim}
    \vspace{-0.5em}
    \caption{Linked list.}
    \label{fig:ir-example-ll}
  \end{subfigure}
\hfill
  \begin{subfigure}[t]{0.45\textwidth}
    \vspace{0pt}
    \begin{verbatim}
Arbor2DArray:
    meta:
        type = "2d_array"
        ...     # optional metadata

    rows = [
        Row(children = [
            Element("A", "1"), Element("B", "2"), ...
        ]),
        Row(children = [
            Element("D", "3"), Element("E", "4"), ...
        ])
        ...     # additional rows of elements
    ]
    \end{verbatim}
    \vspace{-0.5em}
    \caption{Two-dimensional array.}
    \label{fig:ir-example-2arr}
  \end{subfigure}

  \caption{\edit{IRs generated by \arbor for arrays (a) with  examples of generalization to linked lists (b) and two-dimensional arrays (c).}}
  \Description{This figure details the intermediate representation for one-dimensional arrays (a), as well as generalization examples to linked lists (b) and two-dimensional arrays (c). The one-dimensional array and linked list representations are side-by-side above the two-dimensional array representation.}
  \label{fig:ir-example}
\end{figure}


\paragraph{Output} \arbor uses the IR to generates three accessible representations \cut{\textbf{(GP4)}}(Figure~\ref{fig:system-diagram}c)--- tabular, \cut{ representation organized by order and explicit relationships, (2) a} navigable, \cut{representation supporting screen reader navigation and interactive exploration,} and tactile---\cut{preserving spatial and topological structure}and each output is automatically generated from the IR, so\cut{ updates to the model} \edit{any update to the input is} \cut{ are}automatically reflected across all representations. Section ~\ref{sec:arbor-outputs} provides a detailed description of each output representation.

\subsection{Interface}\label{sec:arbor-interface}
\arbor was developed following WCAG 2.1 AA guidelines \cite{wcag21}\cut{,} to ensure accessibility, and tested with the JAWS\footnote{\url{https://www.freedomscientific.com/products/software/jaws/}}, NVDA\footnote{\url{https://www.nvaccess.org/download/}}, and VoiceOver \edit{(macOS) }\footnote{\url{https://support.apple.com/guide/voiceover/welcome/mac}} screen readers to confirm that interactions follow common patterns for screen reader users.

\paragraph{Editor mode} In editor mode, educators can author diagrams using Mermaid or Graphviz DOT syntax while simultaneously previewing the generated outputs. \cut{The interface requires \cut{a}\edit{an} author-defined data structure\cut{ (either an array or \edit{a} binary tree)} so that the translation stage can correctly compile the IR\edit{.}} The live preview ensures that edits are immediately synchronized across all output modalities, \cut{enabling}\edit{allowing} authors to verify both correctness and accessibility. \edit{Authors can also provide titles and descriptions for diagrams, giving students additional context. If an author-provided description is absent, \arbor generates} concise descriptions, satisfying \DR{4} by providing a brief, high-level explanation separate from the data structure representation itself. For example, for the binary tree in Figure ~\ref{fig:tree-outputs}, if no author description is provided, \arbor displays: \textit{``This binary tree contains 6 nodes and 5 edges. The root node is 3.''}

\begin{figure*}
    \centering
    \includegraphics[width=0.99\linewidth]{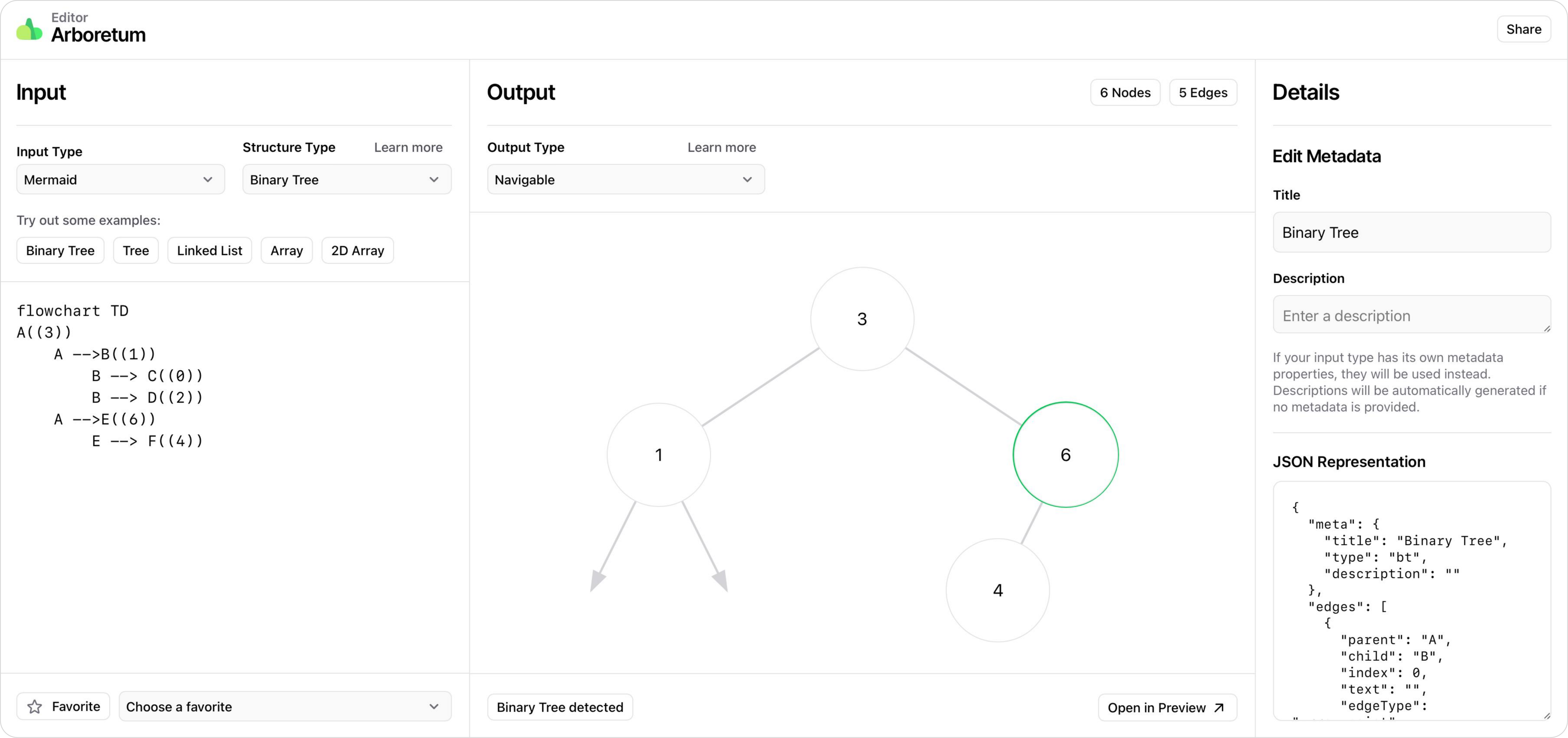}
    \Description{Screenshot of Arboretum's Editor mode. From left to right: an Input pane for writing Mermaid or Graphviz syntax and selecting a data structure; an Output pane showing the Navigable representation with options to switch to the Tabular and Tactile representations; and a Details pane for editing Metadata and viewing the IR in JSON form.}
    \caption{\edit{\arbors Editor mode, which allows educators to write diagram specifications and preview synchronized accessible outputs.\protect\footnotemark[9]}}
    \Description{A screenshot of Arboretum’s Editor mode. In a three pane design, from left to right: the Input pane allows authors to select the input and data structure types to then define the diagram specification; the Output pane allows authors to select the output representation type, displaying the output in the center; and the Details pane allows authors to specify title and description metadata for the diagram as well as view a JSON representation of the generated diagram.}
    \label{fig:arbor-edit-view}
\end{figure*}

\paragraph{Preview mode} In preview mode, students can view the accessible representations of a diagram without interacting with the underlying diagram specification language. The interface \cut{emphasizes}\edit{highlights} the generated representations \cut{with an}\edit{through a} representation-selection dropdown while minimizing distractions from the Editor counterpart, providing a focused space for students to engage with data structures representations.


\vspace{1em}
\noindent
\edit{Across both modes, \arbor supports local favoriting to revisit examples as well as sharing and embedding diagrams via hyperlinks, QR codes, or embedded iFrames. Authors and users can also easily switch between the editor and preview modes to edit or preview the \arbor diagram respectively.}


\cut{Together, these modes provide our accessible data structure representations, which we detail in the following subsection.}




\subsection{Output Representations}\label{sec:arbor-outputs}
\begin{figure}
    \begin{subfigure}[t]{0.4\textwidth}
     \vspace{0pt} 
    \includegraphics[width=0.99\linewidth]{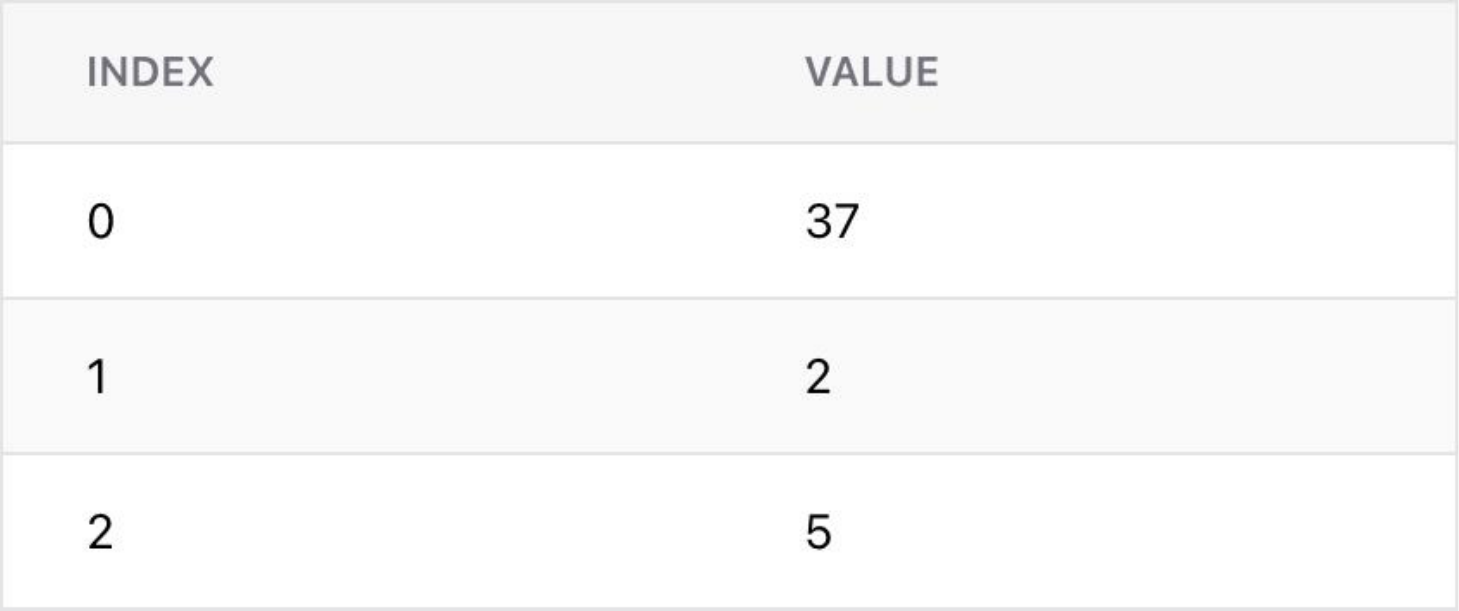}
    \caption{Tabular.}
    \Description{A table listing the index and value pairs of an array.}
    \label{fig:array-outputs-tabular}
    \end{subfigure}
    \begin{subfigure}[t]{0.4\textwidth}
     \vspace{1em} 
    \includegraphics[width=0.99\linewidth]{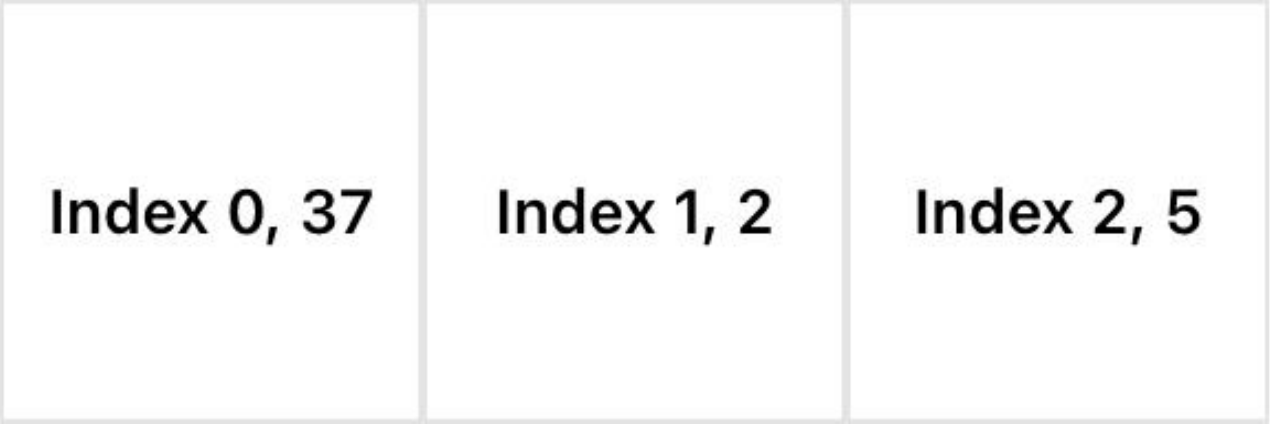}
    \caption{Navigable.}
    \Description{A screen reader-navigable diagram with three adjacent boxes, each containing an index and value.}
    \label{fig:array-outputs-navigable}
    \end{subfigure}
    \begin{subfigure}[t]{0.4\textwidth}
     \vspace{1em} 
    \includegraphics[width=0.99\linewidth]{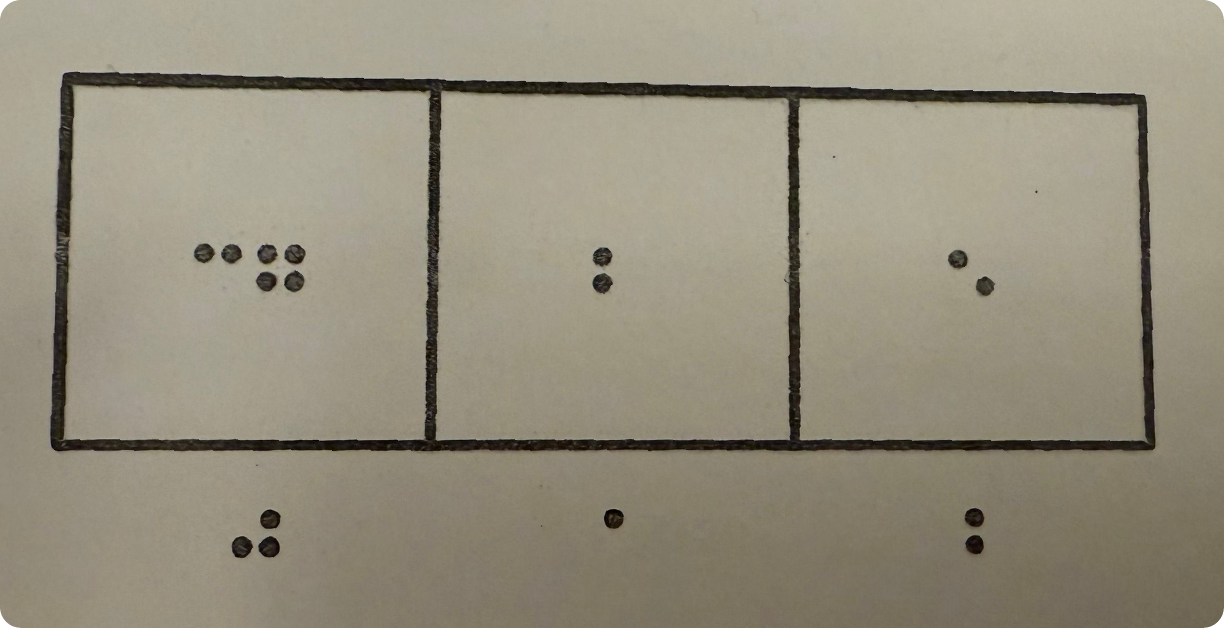}
    \caption{Tactile.}
    \Description{A tactile diagram comprised of three adjacent boxes, each containing a value, with the index below the box using Braille dots.}
    \label{fig:array-outputs-tactile}
    \end{subfigure}
    \Description{A comparison of four different representations for an array data structure containing the values 37, 2, and 5. First, the table lists the index and value pairs. Second, the navigable shows three adjacent boxes, each containing an index and value. Last, the tactile shows three adjacent boxes, each containing a value, with the index below the box using Braille dots.}
    \caption{Representations of an array: (a) a tabular representation of elements' indexes and values, (b) a navigable representation that is screen reader navigable, and (c) a tactile representation with braille \protect\footnotemark.}
    \label{fig:array-outputs}
\end{figure}

\begin{figure}
    \begin{subfigure}[t]{0.4\textwidth}
     \vspace{0pt} 
    \includegraphics[width=0.99\linewidth]{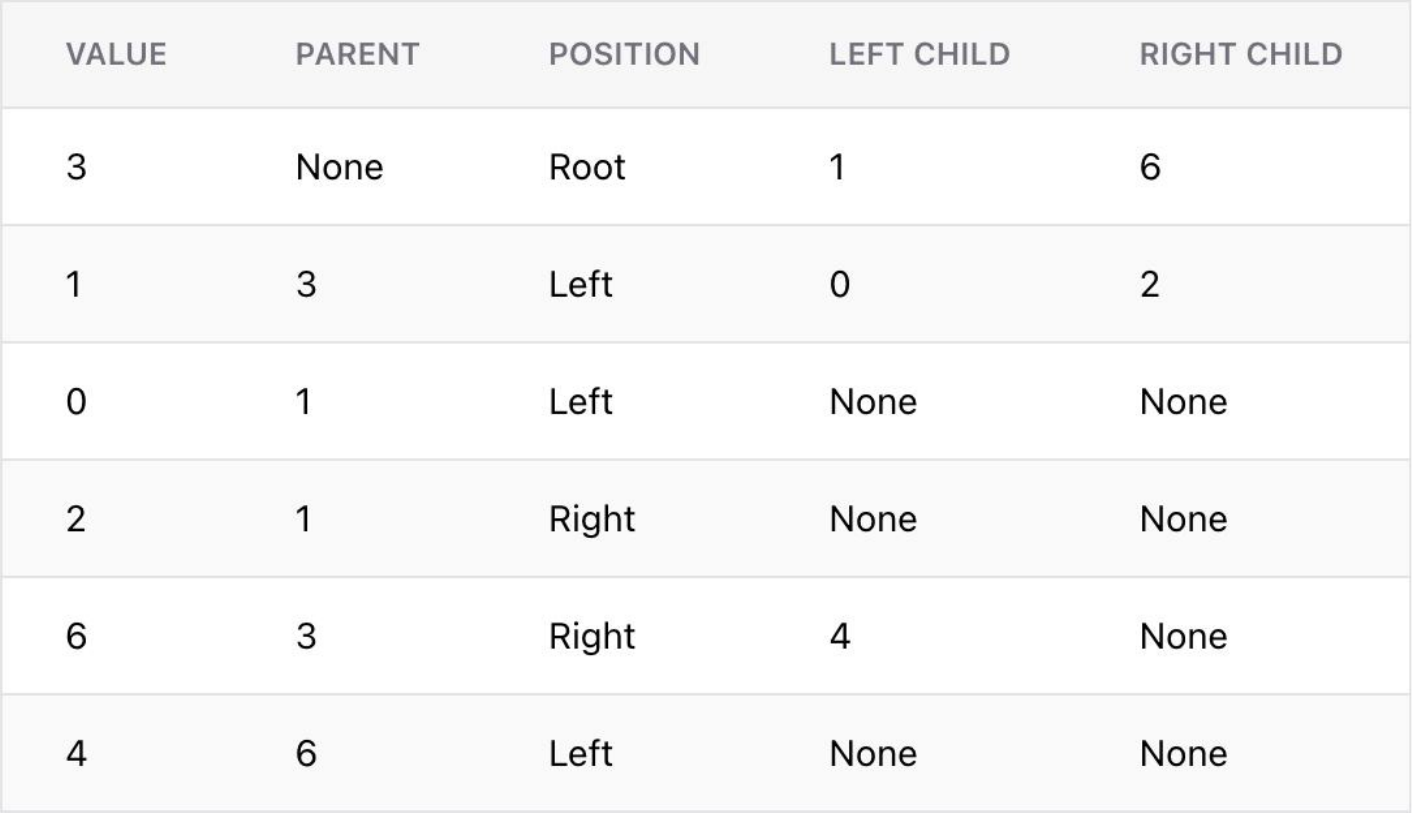}
    \caption{Tabular.}
    \Description{A table lists nodes of the tree, with columns for node "Value", "Parent", "Position", "Left Child", and "Right Child".}
    \label{fig:tree-outputs-tabular}
    \end{subfigure}
    
    \begin{subfigure}[t]{0.4\textwidth}
     \vspace{1em} 
    \includegraphics[width=0.99\linewidth]{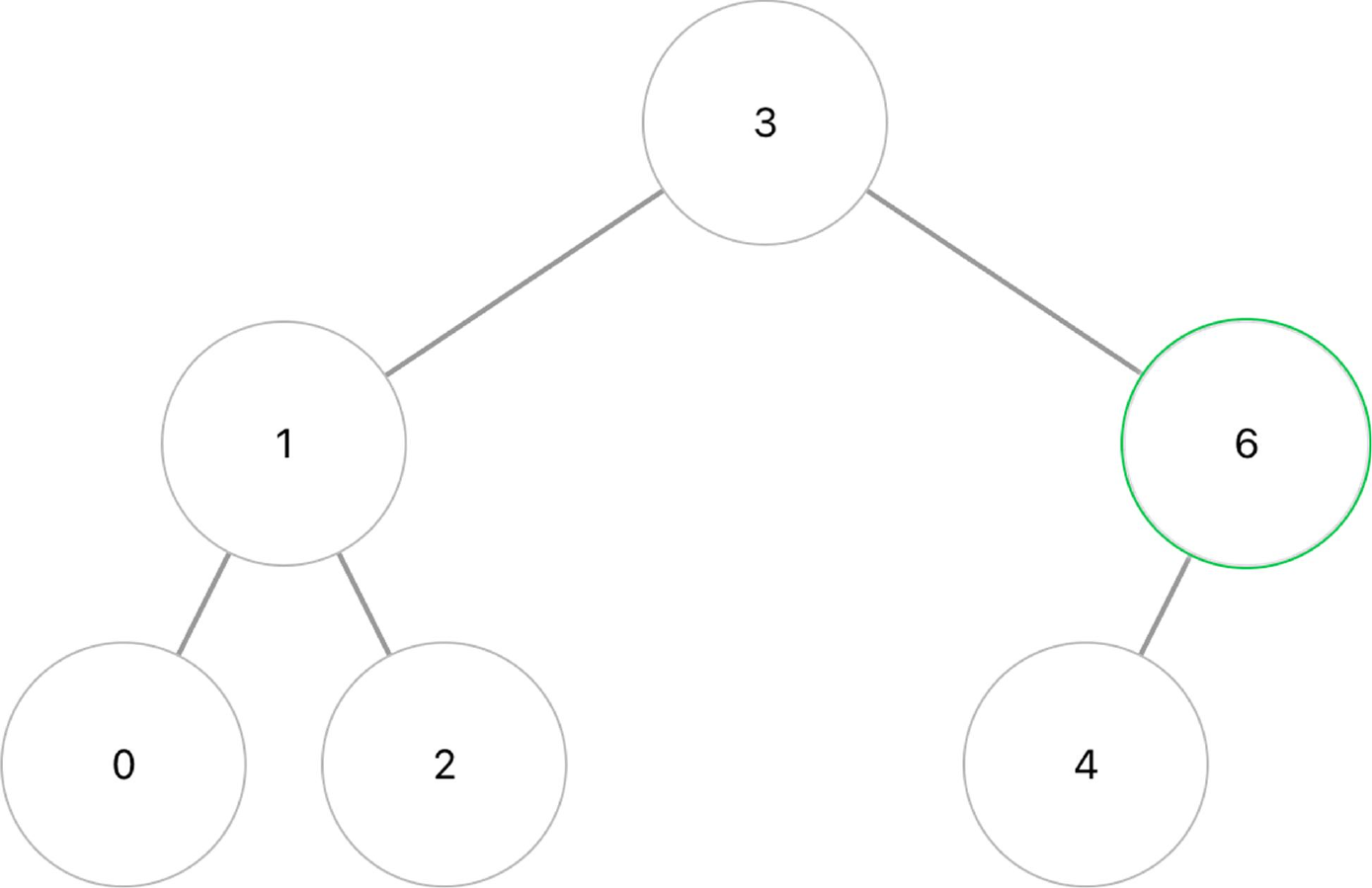}
    \caption{Navigable.}
     \Description{A screen reader-navigable diagram of the tree as a series of line-connected circles, with the values inside.}
    \label{fig:tree-outputs-navigable}
    \end{subfigure}
    
    \begin{subfigure}[t]{0.4\textwidth}
     \vspace{1em} 
    \includegraphics[width=0.99\linewidth]{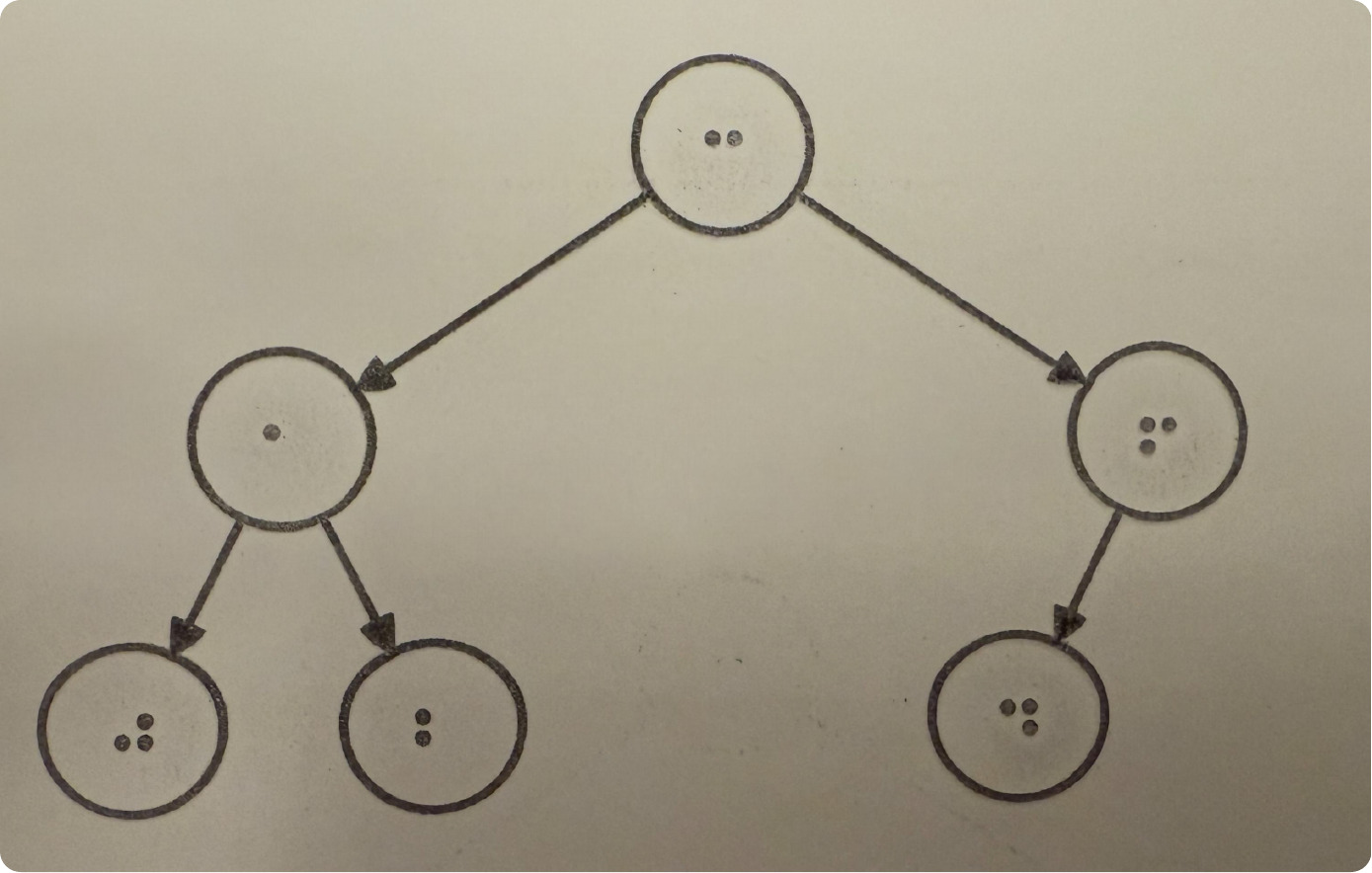}
    \caption{Tactile.}
    \Description{A tactile diagram comprised of a tree with thick black strokes and circular Braille-labeled nodes connected by arrowed edges.}
    \label{fig:tree-outputs-tactile}
    \end{subfigure}
    \Description{A comparison of three different representations for a binary tree data structure. First, the table lists nodes with columns for node "Value", "Parent", "Position", "Left Child", and "Right Child". Second, the navigable shows the tree as a series of line-connected circles, with the values inside. Finally, a tactile displays the same tree structure with thick black strokes and circular Braille-labeled nodes connected by arrowed edges.}
    \caption{Representations of a binary tree: (a) a tabular representation of the parent and children nodes, (b) a navigable representation that is screen reader navigable, and (c) a tactile representation with braille\protect\footnotemark.}
    \label{fig:tree-outputs}
\end{figure}

\arbors pipeline generates three synchronized representations that follow the design requirements identified in Section~\ref{sec:rq}, and are illustrated for arrays (Figure~\ref{fig:array-outputs}) and binary trees (Figure~\ref{fig:tree-outputs}). \cut{supporting complementary pedagogical goals: tables reflect order and linearity, navigable representations enable hierarchical exploration through screen readers, and tactile graphics convey spatial and topographical structure. We illustrate this design with arrays (Figure~\ref{fig:array-outputs}) and binary trees (Figure~\ref{fig:tree-outputs})., highlighting how synchronized outputs provide complementary strengths for exploring data structures.} \edit{Figure ~\ref{fig:og-viz} shows the original visualizations of the array and binary tree provided to sighted students before using \arbor to create alternative, accessible representations.}




 \footnotetext{A screen reader-friendly version of this array diagram is available at: \href{https://g.riteshkanchi.com/chi26-arboretum/array-outputs}{g.riteshkanchi.com/chi26-arboretum/array-outputs}}


\paragraph{Tabular} Tabular representations present \edit{a consistent, unambiguous encoding of }data structures in a table\cut{ structure} format. This modality leverages familiar table navigation patterns and HTML semantics to make structural relationships explicit while supporting\cut{ random access and} efficient nonvisual \edit{access and} exploration. \edit{Given participants’ strong suggestion for tabular formats (\DR{2}), this view also supports \DR{3} by explicitly encoding roles and relationships---such as parent and child---rather than requiring users to infer them from narrative descriptions.} The table encodes the semantics of each data structure: for arrays and lists, rows capture explicit indices and values (Figure~\ref{fig:array-outputs-tabular}); for binary trees, rows capture node attributes such as value, parent/child relationships, and positional role (root, left, right) (Figure~\ref{fig:tree-outputs-tabular}). 


\paragraph{Navigable} While tables expose explicit values and relationships, they do not reflect how learners typically traverse or interact with a data structure. The navigable representation provides an interactive, nonvisual analogue to common navigation patterns (\eg{} \cut{stepping}\edit{iterating} through an array, traversing a tree). This representation maintains the visual and spatial layout \cut{of visual data structure diagrams}\edit{familiar to sighted instructors} but is implemented with screen reader-friendly HTML.  \edit{To satisfy \DR{1}, we grounded the navigation model in established WAI-ARIA interactions---such as list and tree navigation \cite{waiaria}}. Arrays are rendered as horizontally adjacent boxes but are represented as list items containing each element’s index and value (Figure~\ref{fig:array-outputs-navigable}). Binary trees are rendered visually with circular nodes and edges but are represented as nested lists following ARIA tree patterns, enabling users to expand and collapse child nodes for efficient traversal. Following \cut{these}\edit{the WAI-ARIA} conventions preserves visual clarity for sighted learners while offering BVI learners a consistent and accessible interaction model. \edit{\Cref{tab:arbor-tree-nav} summarizes the commands used in the binary tree navigable representation.}

\footnotetext{A screen reader-friendly version of this binary tree diagram is available at: \href{https://g.riteshkanchi.com/chi26-arboretum/tree-outputs}{g.riteshkanchi.com/chi26-arboretum/tree-outputs}}

\paragraph{Tactile} Whereas the tabular emphasizes explicit relationships and navigable supports interactive exploration, the tactile representation provides an explicit spatial dimension. Tactile representations allow learners to physically perceive structure and layout, offering affordances particularly useful for reasoning about hierarchy and positional relationships. In \arbor, the tactile generator produces scalable SVG diagrams \edit{that can be embossed using common tactile-production equipment found in most university assistive technology offices, such as embossers or Swell Form (fuser) machines}. Arrays are represented as adjacent boxes with Braille-labeled values and indices, supporting both sequential and positional reasoning (Figure~\ref{fig:array-outputs-tactile}). \edit{Tactile representations generated by \arbor contain Braille-labeled nodes and arrowheads that mark directed edges for parent–child node connections in a branching layout to preserve hierarchy and directionality (Figure~\ref{fig:tree-outputs-tactile}). Educators can also combine multiple tactile-ready SVGs for a single tactile sheet---for example, showing an original tree and a version with a node removed---by arranging \arbor’s SVG outputs in a slide deck or layout tool.}

\edit{Together, these three representations satisfy \DR{5} by enabling integrated, complementary access: learners can fluidly switch between representations to cross-check relationships, confirm structural details, and form stable mental models.}

\newcolumntype{Y}{>{\raggedright\arraybackslash}X}
\begin{table}[ht]
\centering
\caption{Supported interactions for \arbors Binary Tree Navigable Representation.}
\small

\rowcolors{2}{gray!10}{white}

\begin{tabularx}{\linewidth}{l X}

\toprule
\textbf{Command} & \textbf{Action}\\
\midrule

$\rightarrow \rightarrow $ (Right arrow twice) & Expands the current node and moves to its first child (left child if both exist).  \\
$\leftarrow \leftarrow$ (Left arrow twice) & Collapses the current node and its siblings, then moves to the parent node (no action if no parent). \\
$\uparrow$ (Up arrow) & Move to the prior child; if none, moves to the previous node at the same level. \\
$\downarrow$ (Down arrow) & Move to the next child; if none, moves to the next node at the same level. \\
\bottomrule
\end{tabularx}
\label{tab:arbor-tree-nav}
\end{table}

\subsection{Limitations of \arbor}
While \arbor illustrates a \cut{principled,}\edit{low-barrier to adoption, educator-centered} approach to accessible data structure representations, \cut{and is available as a usable web-based prototype,} several limitations remain. First, only supporting Mermaid and GraphViz DOT may present a learning curve for educators unfamiliar with these languages, and array creation is limited to Mermaid \edit{as GraphViz DOT lacks a commonly adopted convention for block-style diagrams.} Second, screen reader support was evaluated only with JAWS, NVDA, and VoiceOver\edit{ (macOS)}; future work should test Narrator\footnote{\url{https://www.microsoft.com/en-us/windows/tips/narrator}}\edit{,  mobile screen readers such as VoiceOver (iOS, iPadOS) and }TalkBack\footnote{\url{https://support.google.com/accessibility/android/answer/6283677}}, and additional assistive technologies to ensure broader compatibility. Third, node and element labels are length-limited---approximately ten characters visually and three characters in Braille for tactile representations---restricting use of larger values. 

\edit{Tactile output also carries inherent constraints independent of \arbor. Physical production technologies (\eg, swell paper, embossers) impose constraints on sheet size, resolution, and production cost \cite{improving-understanding, prescher2014production}. In our case, a standard Swell Form machine (\$1{,}475) \footnote{\url{https://tinyurl.com/swell-paper-machine}} and a box of 100 sheets of swell paper (\$138) \footnote{\url{https://tinyurl.com/swell-paper-cost}} kept per-page costs low enough for routine instructional use. In addition, tactile media does not scale well to very large structures such as multi-level search trees or arrays with hundreds of elements, where multiple sheets or sequential prints would be required. For these larger or more complex structures, \arbors digital modalities (\eg tabular and navigable) are better suited for nonvisual exploration; however, large tables can themselves become cumbersome to navigate with a screen reader \cite{wang2024table}.}



\cut{In contrast, the tabular and navigable representations avoid these limitations, whereas physical production still requires educators to carefully consider page size and layout in preparing tactile graphics. While \arbor is stable enough to support real use and participant studies, as with any research prototype, some edge cases and bugs may remain, and we encourage community testing and feedback.}

\section{Evaluating Accessible Arrays and Binary Trees with BVI Users}
\label{sec:study}
We conducted a within-subjects study to examine how participants perceive multimodal access to data structures through three nonvisual representations and how these representations affect comprehension, navigation, and confidence. \edit{Our user study was guided by four specific research questions:} \cut{Our user study directly operationalizes our broader research questions into four specific, measurable questions:}

\begin{itemize}
    \item \textbf{RQ-Comprehension:} How do \arbors representations (tabular, navigable, tactile) support \textit{comprehension} of arrays and binary trees?
    \item \textbf{RQ-Navigation} How do participants \textit{navigate} and interact with the different representations?
    \item \textbf{RQ-Confidence:} How does multimodal access (having multiple representations available) support participants' \textit{confidence} in understanding across data structures?
    \item \textbf{RQ-Application:} How do participants \textit{apply }binary search using \arbors representations, and how efficiently do they do so?  
\end{itemize}

\textbf{RQ-Comprehension} and \textbf{RQ-Application} examine \cut{\textbf{RQ1} by testing }whether participants can extract computational information (element location, BST properties, algorithmic steps) from each representation. \textbf{RQ-Navigation} explores the interaction mechanisms that make these properties accessible, while \textbf{RQ-Confidence} \cut{addresses \textbf{RQ2} by measuring} \edit{measures} how multimodal access affects participants' certainty and reasoning strategies across different data structures.

\cut{With our focus on arrays and binary trees, w}\edit{W}e designed each study session to build conceptually from simpler to more complex tasks. Participants first explored arrays, a linear data structure that establishes basic notions of indexing and bi-directional navigation. Next, they explored binary trees, which introduce hierarchical relationships and more complex navigation challenges. Finally, participants applied the binary search algorithm on binary trees, requiring them to integrate structural understanding with algorithmic reasoning. This progression allowed us to examine how \arbors representations supported learning from foundational comprehension through to applied problem-solving. 

\subsection{Participants}
We recruited \edit{8 (5 female,} 3 male) BVI participants through accessibility-focused mailing lists and referrals from organizations serving BVI individuals. Our inclusion criteria included participants that: (1) were at least 18 years old, (2) self-identified as blind or visually impaired, and (3) were comfortable navigating digital content with commonly-used screen readers (\eg{} VoiceOver, JAWS, NVDA). \edit{One additional participant completed the study, but was excluded from analysis because she was not familiar with standard screen reader interactions during the session, despite screening as comfortable with a commonly used screen reader. As a result, her data was not comparable to that of other participants.} Although prior experience with CS was not required, half of respondents (n=4) described having some prior exposure that came through formal education, work, or informal learning. Participants ranged in age from 37–65 (M = 50.8, SD = 12.4), with vision loss occurring at different stages of life. Each participant received a \$40 Amazon gift card as compensation for their time. Table~\ref{tab:eval-demographics} presents participants' demographics. 


\newcolumntype{Y}{>{\raggedright\arraybackslash}X}
\begin{table*}[t]\centering
\centering
\caption{Evaluation Participant demographics.} 
\small

\rowcolors{2}{gray!10}{white}

\begin{tabularx}{\linewidth}{l l l X l X l}

\toprule
\textbf{ID} & \textbf{Gender} & \textbf{Age} & \textbf{Vision Level} & \textbf{Education} & \textbf{CS Experience} & \textbf{Env.}\\
\midrule

P1 & Male   & 43 & Totally blind (no light perception) & Some college & Took a CS class in school; CS-related job; studied CS independently & Remote \\
P2 & Male   & 41 & Totally blind (some light perception) & Bachelors & None  & Remote \\
P3 & Female & 62 & Low vision (some usable vision) & Masters & Studied CS independently & Remote \\
P4 & Female & 65 & Totally blind (some light perception) & Masters & None   & In-Person \\
P5 & Female & 37 & Totally blind (no light perception) & Masters & None  & In-Person \\
P6 & Female & 64 & Low vision (some usable vision) & Masters & CS-related job  & Remote \\
P7  & Male & 57 & Totally blind (no light perception) & Bachelors & Took a CS class in school; CS-related job; studied CS independently & Remote \\
P8  & Female & 37 & Totally blind (some light perception) & Masters & None & Remote \\
\bottomrule
\end{tabularx}
\label{tab:eval-demographics}
\end{table*}

\subsection{Task Design and Structure}
The study design included $6$ tasks (T1-T6, shown in \Cref{tab:tasks}). 
\cut{one of which was excluded from data analysis due to faulty design.} The tasks were chosen to align with our research questions, and originally included two array tasks (T1 and T2) and four Binary Tree tasks (T3-T6). Each task was designed to require interacting with the data structure to answer one or more questions. Participants were free to use whichever representation(s) they found most most helpful, switch between them as needed, and were encouraged to think aloud while working. \edit{For the study, we simplified  \arbors preview mode by removing sharing, local favoriting, and editor mode, allowing participants to focus solely on interacting with the output representations.}

The number of questions varied across tasks. Questions probed concepts such as array indexing (T1), parent–child relationships, leaf identification, and verifying the BST property (T3–T5). Finally, a binary search task (T6) included questions that probed participants' algorithmic reasoning.  

\cut{Participants were scored on task questions.} After each task question, we asked participants to think aloud while answering a more detailed follow up question. 
For example, to answer T4, participants counted the number of leaf nodes in the tree. The follow up question was `what are their values?' Our intent was to understand the role of the representations in their learning. 

\cut{In the case of T2, our process uncovered three warning signals. First, participant answers took less time than any other question (1.1s on average vs 6.14-55.43s on average for the other questions). Next, participants did not describe data structure use in the follow up question. Finally, there was no correlation with \textbf{RQ-Comprehension}, \textbf{RQ-Navigation}, \textbf{RQ-Confidence}, or \textbf{RQ-Application}. Our interpretation is that participants may have answered T2 from memory alone.}






\begin{table*}[t]\centering
\centering
\caption{Task list for the Arboretum user study. 
All specific tasks can be found in Supplemental Materials. \# is \# of Questions}
\small
\rowcolors{2}{gray!10}{white}
\begin{tabularx}{\linewidth}{l l l  X r}
\toprule
\textbf{Task} & \textbf{Category} & \textbf{Structure} & \textbf{Example Question(s)} & \textbf{\#}  \\
\midrule
T1 & Element Location & Array & Is the element 54 in the array? If so, at what index? & 3\\
T2 &Order/Sorting & Array & Is the array sorted in increasing order? & 1 \\
T3 & Parent/Child Identification & Binary Tree & What is the value of the root's right child? & 1\\
T4 & Leaf Node Identification & Binary Tree & How many leaf nodes are in the tree? & 1\\
T5 & BST Property Check & Binary Tree & Do all nodes satisfy the BST property? & 1\\
T6 & Binary Search & Binary Tree & Does the tree contain the numbers 0, 5, 6, or 9? & 3 \\
\bottomrule
\end{tabularx}
\label{tab:tasks}
\end{table*}

\subsection{Procedure}
Sessions were conducted either remotely or in person and lasted approximately 60 minutes. Participants received a consent form and a short demographic survey in advance. At the start of each session, facilitators confirmed participants' background and CS experience, addressed any questions, and ensured both consent and the demographic survey were completed. To enable consistent screen recording of computer interactions, all participants joined a Zoom call, including those attending in person. Once these steps were completed, participants shared their screens so facilitators could observe their interactions.

We employed a within-subjects design in which all participants completed activities for both arrays and binary trees. \edit{All participants completed the same tasks in the same conceptual order. Because our goal was to examine how participants reasoned with full access to all representations—rather than compare performance across modalities—we did not create separate task variants. Instead, participants chose whichever representation(s) they preferred, enabling us to observe authentic strategy selection.}

The three representations---tabular, navigable, tactile---were presented in a counterbalanced order. \edit{Counterbalancing applied only while participants were learning each representation, ensuring that no modality benefited from being taught first while participants were still forming initial mental models of the data structures. After this phase, participants were free to use any representation(s) for the tasks.} \cut{to control for ordering effects and reduce bias from participants encountering the exact same representation first as they learned about each data structure. }

The activity sequence was fixed (arrays $\rightarrow$ binary trees $\rightarrow$ binary search),  \edit{reflecting a common progression from simple linear (\ie{} arrays) to hierarchical structures (\ie{} binary trees) to algorithmic operations over those structures. Each task was administered once per participant to prioritize rich think-aloud reasoning rather than repeated-measures performance comparisons.} \cut{due to the conceptual dependencies between activities. Arrays establish foundational concepts of sequential ordering and index-based access, prerequisite for understanding tree navigation. Binary trees introduce various traversal strategies \cite{data-structures-and-their-algos}, essential for applying binary search algorithms.}



For each activity, we began with a brief introduction of the data structure and its application, followed by the representations\cut{ in their counterbalanced order}. Facilitators guided participants through practice questions using a simple example to illustrate how each representation could be applied to answer structural questions. For remote participants, tactile materials were shipped in advance, and we standardized orientation by cutting the top-left corner of each page and embossing tactile page numbers in the top-right corner.

This introduction allowed participants to see how different representations afforded different strategies. For instance, a leaf node could be identified in the tactile graphic by locating nodes without outgoing edges, in the table by finding rows where left/right child columns were empty, or in the navigable representation by checking which nodes could not be expanded or collapsed. Once participants were comfortable using the representations, they were  provided with the official task (T1-T6) for that data structure.

For arrays, the \cut{practice}\edit{learning} example \cut{contained}\edit{included} three elements, while \cut{the task example}\edit{T1 and T2} \cut{contained}\edit{included} eight elements. For
binary trees, the practice example \edit{was a binary tree of height 3 with 6 nodes, while T3-T5 had a binary tree of height 3 with 7 nodes (a full binary tree).}\cut{used a perfect BST with three levels, while the binary tree task used a non-BST
structure.} \cut{For the binary search activity, the task used a four-level BST to support algorithmic reasoning.} \edit{Prior to T6, participants were taught binary search step-by-step using the practice example in their preferred representation, giving them an opportunity to learn the procedure before applying it independently. The practice tree satisfied the BST property and differed in both structure and values from the T6 tree; T6 had a binary tree of height 4 with 9 nodes that did not satisfy the BST property.}

Participants were free to use any representation(s) they preferred to complete the tasks listed in Table~\ref{tab:tasks}. At the end of each activity, participants completed a short 5-point Likert-scale survey rating their comprehension of the data structure, clarity of the representation, ease of navigation, confidence in their answers, and how well each representation conveyed the underlying structure. After all activities were complete, we conducted a short semi-structured interview to elicit reflections on the usability of each representation, the experience of having multiple representations available, strategies for switching among them, and the transfer of understanding across data structures (\eg{} moving from arrays to binary trees).


\subsection{Data Collection and Analysis}

\edit{We recorded the time needed to answer each question, and the correctness of the answer. In the case of T2, our process uncovered three warning signals. First, participant answers took less time than any other question (2.1s on average vs 6.9-62.5s on average for the other questions). Next, participants did not describe data structure use in the follow up question. Finally, there was no correlation with \textbf{RQ-Comprehension}, \textbf{RQ-Navigation}, \textbf{RQ-Confidence}, or \textbf{RQ-Application}. Our interpretation is that participants may have answered T2 from memory alone. For this reason we excluded T2 from our quantitative analysis. For the remaining tasks, we report descriptive statistics because of}
\cut{Given} our small sample size (n=\cut{7}\edit{8}). \cut{we report descriptive statistics (means and standard deviations) for quantitative measures rather than inferential tests.} This approach aligns with recommendations for small-sample accessibility research \cite{lazar2017}. Our primary insights derive from an inductive thematic analysis of participants' interactions and feedback \cite{braun2006using}, with quantitative measures providing complementary context about performance patterns.

For our descriptive statistics, we used Accuracy of Extracted Information (AEI) as our primary performance measure, following prior work in accessible data representations \cite{hoque23sound,sharif2022voxlens}. AEI is coded as a binary outcome for each task question, with a value of 1 (``accurate'') if the participant answered correctly and 0 (``inaccurate'') otherwise. Overall accuracy for each task was calculated as the proportion of correct responses out of the total number of questions, expressed as a percentage. In addition to AEI, we collected time spent for each task, time spent using each data structure, and  participants' Likert-scale ratings after each activity. For tasks with multiple questions, (\eg{} T1, T6), we calculated the mean accuracy, mean time spent, and standard deviation of time spent across questions within each task.


For qualitative analyses, we analyzed participants' anonymized transcripts and screen recordings. Two authors independently conducted open coding, generating 444 initial codes that captured fine-grained aspects of participants' feedback. Through iterative discussion and constant comparison, these were consolidated into a codebook of 78 focused codes. The authors then collaboratively organized codes into 17 higher-order categories and 5 overarching themes: representation affordances \& preferences, multimodal integration \& strategic use, learning processes \& mental model development,conceptual understanding \& algorithm application\edit{, and accessible education implications}.


\section{Results}
\label{sec:findings}
\cut{Our analysis combines quantitative performance measures with rich qualitative insights from participants' interactions and reflections.} Our results begin with an overview of task performance and subjective ratings to establish the quantitative landscape, then explore five key themes that emerged from our thematic analysis. \cut{: how individual representations afforded different interaction possibilities, how participants developed navigation strategies and combined representations, learning processes and transfer to algorithmic reasoning, and broader accessibility design implications.}

\subsection{Overview of Performance and Preferences}

Participants demonstrated strong performance on most tasks, with accuracy patterns supporting the effectiveness of our representations. As shown in Table~\ref{tab:task-time-perf}, participants achieved perfect accuracy on some tasks (parent/child identification, binary search), while encountering the most difficulty in checking the BST property across all nodes in a tree.

Our array task revealed strong understanding in both performance and timing \edit{(T1, 91.67\% accuracy, M=22.8s, SD=6.9s)}. Participants were effective in finding the location of an element based on the value or index, as well as determining adjacent element values. However, some participants struggled in conditional aggregation (i.e., counting the number of elements greater than 10) across elements \edit{as they needed to maintain additional information (a counter) while iterating through the array.}


%

Binary tree tasks revealed greater complexity in both performance and timing. While participants successfully identified parent-child relationships \edit{(T3, 100\% accuracy)}, \cut{more challenging tasks showed}\edit{there was} decreased performance\cut{:}\edit{ for} leaf node identification \edit{(T4, 87.5\% accuracy)} and BST property verification \edit{(T5, 62.5\% accuracy)}. These complex tasks also required substantially more time \edit{(T4: M=39.8s, SD=39.5; T5: M=62.5s, SD=52.1). BST verification saw the sharpest drop because the task required participants to \edit{keep track of} several structural relationships at once: tracking values, determining left–right positioning, and maintaining depth awareness.}

The binary search application (T6) demonstrated high task accuracy, with participants achieving 100\% accuracy across all search queries in moderate time \edit{(M=35.5s, SD=13.9)}. Our accuracy measure (AEI) reflects whether participants provided correct final answers, not whether they specifically employed binary search algorithms to reach those answers. \cut{We therefore turn to Section~\ref{algo}, where we }\edit{Section~\ref{algo} further} examine\edit{s} participants' reasoning processes and strategies in detail to show how binary search was carried out in practice.

Confidence measures showed that participants felt highly confident with array understanding and navigation (M=4.88) but somewhat less confident throughout binary tree tasks \edit{(M=4.25-4.50)}. Participants expressed strong confidence in their binary search understanding \edit{(M=4.25-4.63)}. While confidence ratings alone do not equate to learning, when considered alongside participants’ high task accuracy, they suggest that participants developed effective strategies for applying binary search.

While participants were required to use all representations in the initial learning phase, they could choose to use either a single representation or multiple representations when completing the tasks. With this option available, participants strongly preferred the tactile graphics over other representations throughout activities. Three participants used tactile representations exclusively for all tasks. Another three relied on tactile alone for at least three out of five tasks. Only one participant preferred a modality other than tactile: P6, who used tactile with navigable in T4, tactile alone in T5, and navigable alone in the remaining tasks. Overall, across the \cut{35}\edit{40} task decisions (\cut{7}\edit{8} participants x 5 tasks), participants selected tactile \cut{27}\edit{31} times, combined modalities \cut{5}\edit{6} times (tabular + navigable + tactile in one case, tabular + tactile in three cases, and navigable + tactile in \cut{one case}\edit{two cases}), and non-tactile modalities only 3 times.

Participants consistently rated tactile graphics highest across all activities (M=5.0 for all tactile measures in ~\Cref{tab:ratings}), indicating strong overall preference for this representation. In contrast, ratings for tabular and navigable representations varied by task complexity. For arrays, both tabular and navigable representations received moderate ratings \edit{(M=4.25)}, but for binary trees, these ratings dropped substantially \edit{(tabular: M=3.38, SD=1.51; navigable: M=3.38, SD=0.52)}. 


\begin{table*}[t]\centering
\centering
\caption{Quantitative measures of task performance. The row shown in \textit{\textcolor{red}{italics and red}} is marked ``T2'' because it was excluded from analysis.}
\small
\rowcolors{2}{gray!10}{white}
\begin{tabularx}{\linewidth}{X p{2.5cm} p{2.5cm} p{2.5cm}}
\toprule
\textbf{Task (Category)} & \textbf{Accuracy} & \textbf{Avg. Time (sec)} & \textbf{Std. Dev (sec)} \\
\midrule
T1 (Element Location) & 91.67\% & 22.79 & 6.89 \\
\textit{\textcolor{red}{T2 (Order/Sorting)}} & \textit{\textcolor{red}{100\%}} & \textit{\textcolor{red}{2.13}} & \textit{\textcolor{red}{1.55}} \\
T3 (Parent/Child Identification) & 100.00\% & 6.88 & 3.52\\ 
T4 (Leaf Node Identification) & 87.50\% & 39.75 & 39.52\\ 
T5 (BST Property Check)& 62.50\% & 62.50 & 52.09 \\ 
T6 (Binary Search) & 100.00\% & 35.54 & 13.87\\ 
\bottomrule
\end{tabularx}
    \label{tab:task-time-perf}
\end{table*}

\begin{table*}[t]\centering
\centering
\caption{Likert-scale ratings (1-5 scale). Top scoring items begin with a * and scores are shown in bold (\textbf{5}). Items with scores below 4 or with  standard deviation $>=1.00$ begin with a - and scores are shown in \textcolor{red}{\textit{italic red}}}

\small
\rowcolors{2}{gray!10}{white}
\begin{tabularx}{\linewidth}{X p{2cm} p{1.5cm}}
\toprule
\textbf{Question} & \textbf{Avg. Rating} & \textbf{Std. Dev} \\
\midrule
I understand the overall structure of a array & 4.88 & 0.35  \\
I feel confident navigating an array using the representation(s) provided. & 4.88 & 0.35 \\
*The format(s) made it easy to locate specific nodes or values in the array & \textbf{5.00} & 0.00 \\
-The table made it easy to understand the array & 4.25 & \textcolor{red}{\textit{1.16}}  \\
The keyboard-friendly rep made it easy to understand the array & 4.25 & 0.71  \\
*The tactile graphic made it easy to understand the array & \textbf{5.00 }& 0.00  \\
I understand the overall structure of a binary tree & 4.50 & 0.93  \\
I feel confident navigating a binary tree using the representation(s) provided  & 4.25 & 0.89  \\
The format(s) made it easy to locate specific nodes or values in the binary tree & 4.50 & 0.53  \\
-The table made it easy to understand the binary tree. & \textcolor{red}{\textit{3.38}} & \textcolor{red}{\textit{1.51}}  \\
-The keyboard-friendly rep made it easy to understand the binary tree. & \textcolor{red}{\textit{3.38}} & 0.52  \\
*The tactile graphic made it easy to understand the binary tree. & \textbf{5.00} & 0.00  \\
I understood how binary search works after today's activities. & 4.63  & 0.52  \\
I could use binary search to solve a different problem or search a different tree if asked. & 4.25 & 0.89  \\
I would feel confident completing a new binary search task on my own. & 4.13 & 0.83  \\
-The table made it easy to complete binary search. & \textcolor{red}{\textit{2.75}} & \textcolor{red}{\textit{1.04}}  \\
-The keyboard-friendly rep made it easy to complete binary search & \textcolor{red}{\textit{3.50}} & 0.53  \\
*The tactile graphic made it easy to complete binary search & \textbf{5.00} & 0.00  \\

\bottomrule
\end{tabularx}
    \label{tab:ratings}
\end{table*}

\subsection{Representation Preferences \& Affordances (RQ-Comprehension, RQ-Navigation)}


The majority of participants preferred the tactile representations for all tasks. 
\cut{They sometimes}\edit{Some} characterize\edit{d} their use of the (preferred) tactile representation as ``cheating'' or ``shortcuts'' because the representation made the tasks very easy to accomplish.\cut{That said, } \edit{Participants recognized the value of the representations, as }\cut{E}\edit{e}ach \cut{representation} offered distinct affordances while imposing specific constraints on how participants could access and manipulate data structure information. \cut{Participants recognized the value of each representation.} As P4 put it, ``I would be thrilled to only even have one.''

The enthusiasm for the tactile representation reflected the intuitiveness of tactile access, across all data structures, especially for understanding structural and semantic details. As P1 explained, ``the tactile was by far the best for the tree representation, it just made it intuitive to see or feel \ldots  how things were laid out.'' Participants emphasized that tactile graphics supported a fundamentally different kind of spatial understanding than digital alternatives. P6 described this distinction directly: ``The answer to that is actually 5 because there's the spatial representation that's not just listening.'' \edit{P8 similarly described their preference for the tactile representation: ``I definitely did like the tactile representations. I feel like I was able to really follow along \ldots you know, literally at my fingertips.''} Others reinforced this enthusiasm with emphatic ratings, such as P2's declaration ``Like a 10. Strongly agree'' and P4's request to go beyond the scale: ``Can I do a 5 1/2? I'll always go for tactile if it's available.''
Tactile graphics also demonstrated efficiency in navigation. Participants described being able to move fluidly across the representation using edge-following strategies and direct pathfinding, in contrast to slower, sequential digital approaches. While our navigable and tabular designs reduced some of the linearity of traditional ALT text, participants noted that they could not match the immediacy of tactile spatial access. P6 highlighted this difference bluntly: ``This is where trying to solve it digitally is a stupid idea because it takes longer.'' This efficiency stemmed from the ability of tactile diagrams to preserve spatial layout and afford immediate, holistic access to structure.

Most significantly, tactile representations supported structural reasoning by making broader structural relationships perceptible at a glance (or touch). Rather than verifying properties node by node, participants could scan spatial layouts—such as left-to-right ordering of values—and detect if the BST property held. As P1 put it, ``\ldots  the details, kind of, popped when I was going through it.'' This sense of details ``popping'' illustrates how maintaining semantic relationships in a tactile representation allowed participants to grasp structural properties rapidly and with confidence, directly linking tactile affordances to reasoning processes.

In contrast, performance of navigable representations was strongly connected to the specific data structure being explored. For example, some participants recognized the familiarity of the interaction model for tree structures 
but still found it difficult to understand hierarchical levels. P7 described this difficulty: ``I have to figure out which one is the hierarchy first basically. Yeah. So that is the most difficult thing I work on.'' These struggles reflect the fundamental challenge of representing vertical hierarchy through screen reader-based tree interactions, where level awareness and local position is not always apparent.
Participants appreciated the navigable list over the navigable tree because it mirrored the conceptual structure of arrays. P6 emphasized the scalability of this approach, explaining that ``for definitely a larger number of elements one wants to have the keyboard friendly.'' However, they still preferred tactile representations and viewed the navigable list as a workable compromise rather than an optimal solution for supporting complex reasoning. P2's mixed assessment reflected this balance: ``I'm going to [rate the list] 3 cause it's like I can see the value of it. But I don't think it's necessary [when the tactile format is available]''

The tabular representation was similarly helpful for arrays, as P4 describes: ``The table to me is the clearest because I seem to understand things that way more easily.'' P1 describes how tabular representations preserved linear order while supporting multiple access pathways: ``I can just follow through on either the values or the indices.'' Tables also helped participants organize information mentally. P2 explained, ``In my head it keeps it organized better. So I can picture it.'' 
However, tables showed limitations with mapping the concept of a binary tree. 
 P1 articulated this challenge, ``I felt like the tabular version of the tree\ldots the concepts didn't map onto it very well. Because it really is just a 2D array, it doesn't behave like a tree.''However, P1 still described them as acceptable fallback options: ``For the binary tree \ldots . If you don't have a choice and have to just do it in a linear format, the table did the job.'' P2 emphasized their conditional utility: ``I think the table is of value \ldots  once I would get familiar with the concepts and what I was looking for.'' Together, these reflections suggest that tables were well-suited for arrays but functioned only as partial substitutes for binary trees.

\subsection{Multimodal Integration \& Strategic Use (RQ-Confidence, RQ-Navigation)}
While tactile representations overwhelmingly dominated task time (\edit{seven}\cut{six} of \edit{eight}\cut{seven} participants used them for over 89\% of recorded time), participants did not rely on tactile alone for all purposes. Three relied exclusively on the tactile format, while \edit{five} supplemented it with at least one other representation.
During the introduction of each data structure, participants often made comments such as ``ahh that makes sense,'' reflecting how exploring all three representations deepened their understanding. Some participants noticed how once they used tactile, they understood the tabular and navigable better. 
Participants also switched between representations to support verification and discovered context-specific advantages that supported efficiency and comprehension. For examples, P3 first checked the tactile representation by scanning nodes for a value, then verified their answer in the tabular representation: "\ldots I started with the tactile graphic, and I went by each [node]\ldots for each one of [the nodes], I would check to find the value. And then I went back to the table and looked for the value."

\paragraph{Multi-modal verification}
During guided learning, and occasionally task completion, participants valued alternative representations as a way to verify their answers.
When being introduced to the binary tree navigable representation, P1 cross-checked details using the tactile representation. For example, they misidentified a node as the left child in the navigable representation, then reached over to the tactile representations to confirm: ``I'm reaching over to check [tactile]\ldots I want to check on the, uh\ldots  okay, no. 2 is the right child of one.'' 
P3 also expressed interest while learning the binary tree navigable representation to verify with the tactile graphic: "So what you might have seen. But after I looked at using the keyboard friendly, I then wanted to do a double check with the tactile graphic." 
We saw this strategy throughout the session, where participants would switch to an alternative representation to formalize and verify before confirming their answer. 

\paragraph{Synergistic Effects}

As P6 described, having access to multiple representations was especially important during the introduction and practice phase, when participants were still learning how to interpret them: ``The graphical table one or Braille one, tactile one and then having the tree one and then OK, here's your table. Here's what to listen for. I would use those as like building blocks. Oh my God. It was essential.''
Moving between representations reinforced participants’ understanding of both the data structures and the other modalities. As P2 explained, ``So like I since they were building on each other and if I only had access to one [representation], once I've had the terminology down and what I was looking for, I think that either of these three would work.''
Access to multiple representations worked synergistically throughout the session, helping participants build data structure understanding.


\paragraph{Context-Dependent Use}
A few participants, such as P5, saw switching between representations as data structure-specific; modality switching was beneficial in learning arrays, but not in binary trees: ``OK, switching helped with the arrays, but then with the tree just confused me.''  P1 also noted that their preference might depend on what device they used: ``if I was on a phone, I might prefer the list. It's just a simpler thing for the array, at least.\ldots versus the table, you know, if I\ldots  If my finger's a quarter inch off, it's gonna be saying that value instead of the index, or something like that.''
These reflections underscore the need for accessible representation design to be cognizant of how individuals might use them in varying contexts.  

\subsection{Learning Processes \& Mental Model Development (RQ-Comprehension, RQ-Application)}
Participants developing  understanding of  data structures revealed complex cognitive processes involving mental model construction, strategic adaptation to barriers, and sophisticated metacognitive awareness. This theme examines the mechanisms through which participants overcame representational challenges and built understanding across data structures. These approaches revealed how participants actively construct meaning when working with accessible representations.
\paragraph{Mental Model Construction}
Participants developed mental models of data structures by combining multiple accessible representations. Mental model construction was most effective when participants integrated insights across representations. P2 described this synergistic process, highlighting the critical importance of the tactile representation: ``I felt like maybe the table allowed me to have a little bit of a picture. And then like it was just great to have the picture in the tactile form. So that just kind of really put it together.'' \edit{Participants emphasized that tactile graphics provided an initial anchor that strengthened their understanding across modalities. As P8 described, if they began with ``either the table or the keyboard representation,'' the tactile version ``kind of amplified my understanding… put an image in my head… so I could follow along,'' ultimately showing how the tactile representation made the digital representations easier to interpret.} This sequential building process, where initial understanding from one representation provided foundation for deeper comprehension through another, demonstrates how multimodal access supports mental model development. 
P3 characterized this as having ``a smaller way of presenting the picture, therefore you had something to build upon,'' highlighting how each representation contributed scaffolding for more complete understanding. 

\paragraph{Analogical Reasoning} Participants often connected data structures to familiar real-world experiences to make abstract concepts concrete. P1 used economic metaphors to understand algorithmic efficiency and the value of the binary search algorithm: ``If you imagine those as safety deposit boxes, and you actually have to open them, and it takes some cost to opening up each one\ldots  the binary tree is clearly better.'' P4 drew on everyday technology to situate their understanding of the tree navigation, reminding themselves: ``Wait a minute. This is just like you're in File Explorer.'' P3 described hierarchical levels in generational terms: "I'm thinking the grandparent\ldots the mom, and then the child underneath." These analogies revealed how participants actively connected abstract computational concepts to familiar  experiences and structures, making them more intuitive to navigate. 

\paragraph{Metacognitive strategies} Metacognitive strategies proved central to participants' success, involving both active self-monitoring and strategic management of their own cognitive resources. P2 demonstrated explicit reasoning checks: ``So that would meet that [BST property]. Would the only one on the left not meet [the BST property]. That would be the one that goes from [node] 2 to [node] 3. Am I thinking of that correctly or not?'' 
In addition, participants showed metacognitive awareness of their own memory limitations, as described by P4:
``My brain works\ldots  I can remember a column-line correlation better than I can remember as I'm moving through something, better than I can remember the other type of correlation.''  Beyond these moment-to-moment checks, participants also engaged in metacognitive reflection about whether their use of a representation aligned with the intended learning objectives.  ``If I'm using this table like this, then I'm really just looking at it as an array. I'm not using the tree\ldots  I feel like that spoils the objective of learning about it as a tree.''  This illustrates how participants not only monitored their reasoning but also evaluated whether their strategy preserved the conceptual integrity of the task.

\subsection{Conceptual Understanding \& Algorithm Application} \label{algo}
Across activities, participants demonstrated varying degrees of conceptual understanding and algorithm application. Their approaches highlighted how prior experience, representational affordances, and terminology shaped both comprehension and problem-solving.

\paragraph{Impact of Prior CS Experience on Understanding}
Participants’ backgrounds in computer science shaped how they approached the representations. While participants' prior computing experience provided some scaffolding—as P7 noted, ``I'm already familiar with array structure because I work on computers''—this familiarity didn't automatically transfer to representation-specific skills. In fact, P7 reflected on how ease of understanding might differ for those without such exposure: ``So people who would not be familiar with this array structure\ldots  I'm not sure how easy it would be, but you know, at least to me it is easy.'' While no participants were previously familiar with binary trees, all were familiar with trees as a concept and as a navigation pattern, having encountered them in screen reader contexts.
Terminology gaps also impacted understanding, as P5 noted: ``I don't know how to explain the relationship between the elements, like the first column header.'' These were particularly salient when participants had partial understanding but lacked precise language to describe relationships.

\paragraph{Semantics}

Participants developed strategies for identifying and understanding semantic roles across different data structures. When working with tree structures, participants demonstrated clear comprehension of hierarchical relationships, as P3 explained: ``On the tactile format, having the representation with the actual lines that showed the parent to child, and it delineated those.''
Identifying leaf nodes illustrated how participants applied semantic concepts in problem-solving, as shown by P2: ``There's only three [leaf nodes]. Because I don't have any lines going from there.'' This conceptual grasp showed value during algorithmic reasoning, as evidenced when P6 performed binary search: ``We've got 3. So 5 is greater than 3. So we gotta go down the right one and then we're up at 6. So 5 is less than six. It's going to be down here. Instead of four, it's a leaf, so there's no 5.'' This example shows how semantic understanding of leaf nodes informed decision-making.
However, participants sometimes encountered conceptual challenges when working to establish semantic clarity. P1 noted how certain formats required additional cognitive effort to distinguish semantic roles: "For the list, just having each one of them be a separate entity, like both the index and the value, I didn't think it was as obvious." These moments reveal the active cognitive work participants engaged in to construct semantic understanding across different representations.

\paragraph{Structural Understanding}
Conceptual understanding also depended on participants' ability to reason about structure—how elements were positioned and connected in ways that supported algorithmic procedures. As GP3 emphasizes, this requires access to topological properties such as order, hierarchy, adjacency, and connectivity. 
For example, 
P4 reflected on how tree traversal required awareness of depth: ``\ldots you're down here somewhere at like the fourth level in the tree, and somebody's looking for something that's over here in the third level of the tree, not on a different branch\ldots '' Tactile cues helped with this, as P1 explained: ``Alright, I'm on 5\ldots  6\ldots  ought to be down and to the right, and it is. And one step took me there.'' \cut{Participants reasoned about levels, directions, and branching in tree structures}
Reasoning also involved reconciling different formats. For example, P1 expected arrays to be laid out vertically, yet
tactile versions were often horizontal:
``The tactile one is horizontal, whereas\ldots  the ones on the screen are vertical. I don't know if that's actually how they're visually laid out.''\cut{P7 extended this reasoning by imagining how arrays might also be represented: 
Such comments highlight how participants were 
 actively working to map their expectations onto the presented layout and to conceptualize orientation in  meaningful terms.}

\paragraph{Algorithmic Reasoning}

Participants demonstrated algorithmic reasoning across representations, applying comparative rules and checking structural constraints. 
P3 explicitly checked BST properties: ``So let me\ldots  before we do that, it has to have. The left child is going to be smaller than the parent. And if the right\ldots  if there is a right child, that child's going to be larger than the parent.'' Both novice and experienced participants executed binary search algorithms through systematic rule application. P7 demonstrated methodical traversal logic: ``Zero is less than 5, so go to the left. Node 0 is less than three. Go to the left, node 0 is less than two. Go to the left\ldots  and zero is less than one, but there's no left node there after that.'' Such careful narrations show how participants internalized the comparative logic of binary search while maintaining awareness of structural constraints like leaf nodes. Verbalization also served as an important strategy for scaffolding reasoning and checking errors. 

Participants further developed nuanced awareness of how different representations shaped algorithmic reasoning. Some formats naturally encouraged systematic approaches, as P1 observed: ``I did, at least with the electronic version of the tree form\ldots  it does kind of push me to use the algorithm so I don't just end up reading every value.'' Others recognized when representational constraints conflicted with algorithmic logic. P7 noted: ``But if I wanna apply binary search, tree view wouldn't work very well because you have to figure which one is left node, which one is right node\ldots  I go back and forth and it keeps interrupting my workflow.''

\subsection{Accessible Education Implications}
\label{sec:acc-edu-impl}

Access to high-quality instruction and accessible materials can fundamentally shape learning trajectories, while persistent systemic barriers  limit equitable participation in computing education. Although we did not explicitly ask participants about their broader educational histories, this theme emerged through our thematic analysis. Participants reflected on the role that accessible representations played—or could have played—in shaping their educational experiences and highlighted the potential impact of \arbor. 


Participants had varied access to tactile graphics in the context of CS education. P1 recalled that: ``Having good computer science instruction\ldots  was so valuable to keeping me interested\ldots I was very lucky in high school computer science, I did have tactile graphics to represent these concepts in my textbook.'' This combination of effective pedagogy and accessible materials sparked sustained interest and self-directed learning that extended far beyond formal coursework.
Other participants reflected on how their educational experiences might have differed with better access to accessible representations. 
P4 expressed regret about missed opportunities: ``I wish I'd had this kind of stuff when I took that one class in computer programming.''
P1 connected this need to self-directed learning: ``My training past high school\ldots  was self-taught, and so having these representations in some of those textbooks would have been great.'' \edit{These experiences shaped long-term engagement, enabling self-directed learning that extended beyond formal instruction. Participants with this prior exposure also articulated a clear sense of tactile’s limits: tactile graphics were foundational for understanding structure, but often arrived slowly or were impractical to produce at scale. As P1 explained, they would still ``prefer one of the electronic versions over waiting… to get a hard copy,'' highlighting a pragmatic shift toward digital formats once a tactile mental model was firmly in place.}

These individual experiences reflect broader institutional challenges. P6 described how educators regularly encounter these gaps: ``This one programmer list I was on, one of the things that happens is they'll be a teacher who's like, I've got a blind student in my class. How do I do data structures?'' Even for  well-intentioned instructors, there is a lack of appropriate resources and training. There are also logistical challenges of relying on tactile graphics, as P1 described: ``I think people are gonna be\ldots  in a lot of cases, stuck with the electronic one if they're remote and they don't have the tools or the agency they're working with doesn't have the resources to mail things.'' \edit{Participants who primarily used magnification software or digital tools similarly emphasized that electronic formats were often more reliable in day-to-day coursework, reinforcing the need for accessible digital modalities alongside tactile ones.}

P6 highlighted that, ``\ldots not everybody has access to like a Braille display or you know [other assistive technologies].'' Also, even when technologies were available, participants noted that screen reader settings and conventions sometimes complicated access. P1 explained, ``I have mine in advanced mode, so it only says the number. For the level, it doesn't actually say level.'' In this case, P1 was describing how customized screen reader verbosity settings change the output: instead of announcing “level 2,” the screen reader only announced “2.” While efficient for experienced users, this shorthand could obscure structural cues (e.g., hierarchy depth in a tree) that are critical for navigating educational representations.


Participants also noted that accessibility barriers extended beyond blindness, reflecting broader educational challenges. One participant with a hearing disability emphasized the limits of relying solely on auditory information: ``Because of the hearing disability, solely relying on speech is not necessarily giving me the most accurate information, so having the tactile again gives me greater independence and ability to manipulate the information.'' Similarly, low-vision participants described how magnification tools often created inefficiencies: ``It's more efficient to not work visually, 'cause at my level of magnification it's off the right side of the screen'' (P6). Together, these accounts highlight the importance of designing representations that are not only conceptually clear but also broadly usable across disabilities and contexts.

\section{Discussion}
Traditional accessibility approaches treat data structure diagrams as visual artifacts to be described, rather than as structured information whose semantics---relationships, ordering, and roles---must be made directly accessible. As a result, standards such as WCAG’s ``complex graphics'' guidance \cite{wcag21} often reinforce surface-level translations of visual diagrams instead of exposing the computational properties that nonvisual learners rely on. \cut{Our work challenges this model by showing accessible representations that preserve the computational properties of data structures can support BVI users to understand data structures.} 

\edit{We discuss our findings into two areas. First, we connect our findings to the four foundational principles of WCAG---that a document is \textit{Perceivable}, \textit{Operable}, \textit{Understandable}, and \textit{Robust}. In doing so, we formalize what accessible, structure-first representations must provide. Second, we discuss  implications for scaling accessible diagram systems, focusing on how educators can generate, adapt, and integrate representations in practice.}

\subsection{Design Principles}

\edit{In this section, we consolidate the five design requirements (\DR{1}– \DR{5}) into four broader design principles that define what structure-aligned accessibility demands. These principles build on WCAG’s perceivable, operable, understandable, and robust (POUR) framework \cite{wcag21}, but extend it to contexts where the underlying structure---not the visual layout---is the primary object of reasoning. 
These principles offer a foundation for accessible representations across data structures and other diagram types.} 


\paragraph{DP1. Perceivable - Explicit Semantic Roles and Relationships} The \textit{Perceivable} principle from the WCAG framework requires that all relevant information be present in a modality that users can receive \cite{wcag21}. Prior accessibility work shows that explicit semantic structure supports efficient nonvisual navigation \cite{fernandes2006transcoding,haque2022grid}. However, past work is premised on an assumption that an element’s role, grouping, and position are fixed---an assumption that breaks down in many diagrams. Even when these properties are stable, their interpretation is often ambiguous without additional semantic context, as a single element can occupy multiple functional roles at once. For example, a node may simultaneously be the left child of its parent, the parent of its own subtree, and a leaf candidate, with these identities shifting as the structure is traversed or modified.  When semantic roles are multi-role and dynamic, relationships are often implicit, multi-functional, and dependent on spatial context. 

\edit{Our Wizard-of-Oz study indicated that no single representation provided enough information for participants to determine these relationships on its own (\DR{3}). Participants faced difficulties identifying ``which nodes were attached to which edges,'' reflecting that the representations presented the elements but did not always make their structural roles clear.} 
\edit{Our second study showed a similar pattern. Within verifying the BST property, participants often had to reconsider a node's role at each step to determine whether it satisfied the property rules. When the representation listed a node’s parent, children, or value explicitly participants used this information to guide their reasoning through analogies that also reflected dynamic roles, such as describing a node as a ``grandparent, mom, and child underneath.''}


\paragraph{DP2. Operable - Structure-Aligned Navigation} 
\edit{Navigation is central to accessible content consumption. WCAG’s \textit{Operable} principle states that users must be able to navigate using a their preferred access technology \cite{wcag21}. Efficient nonvisual navigation benefits from familiar interaction patterns (\DR{1}), and prior research has demonstrated that screen reader navigation improves efficiency and orientation within coding environments \cite{structjumper,potluri2018codetalk,Schanzer:2019:AccessibleAST} and accessing visualizations \cite{thompson-chartreader,blanco2022olli,zong2022rich}. However, screen reader interactions typically rely on tree-based navigation models \cite{blanco2022olli}, which are often incompatible with relational, spatial, diagrammatic, or geographic structures \cite{elavsky-data-nav,benthic2025mei}. 
When underlying structure diverges from a simple tree, familiar interactions such as WAI-ARIA tree navigation, break down---leaving many visualization types unsupported \cite{elavsky-data-nav}. This mismatch has motivated the development of new interaction paradigms for graphs and networks, such as Benthic \cite{benthic2025mei}, that support navigation across both hierarchical and adjacent relationships.}

\edit{Although accessible visualization and interface designers often begin with standard interactions set by WAI-ARIA, these interaction patterns do not always correspond to the structural relationships that data structures express. Across both studies, participants encountered cases where navigation behavior implied one structural meaning while the diagram required another. For example, the right-arrow key, conventionally interpreted as ``move right'' or ``expand'', navigated to a node’s left child due to the collapse/expand interactions in trees. Participants described this as confusing because interaction semantics (e.g., arrow keys as spatial movement) conflicted with the intended traversal semantics.}
\edit{
For data structures with multiple meaningful traversal orders, a single navigation paradigm is unlikely to support all reasoning tasks. A structure-aligned approach should therefore expose more than one traversal logic when appropriate. For instance, enabling a user to move through a tree using depth-first or breadth-first logic, depending on the reasoning task at hand. }


\paragraph{DP3. Understandable - Supporting Structural Reasoning} 
\edit{WCAG’s \textit{Understandable} principle  emphasizes clear and comprehensible content. Prior work on auditory information seeking \cite{zhao2004sonification} and code navigation \cite{structjumper, potluri2018codetalk, Schanzer:2019:AccessibleAST} shows that nonvisual learners rely on multiple access strategies---gists, overviews, orientation cues, and contextual retrieval---to keep users oriented and reason effectively. This work also identifies a recurring tradeoff: interfaces that prioritize faster movement can make it harder to perceive underlying structural relationships, whereas more structured designs may slow navigation but clarify how elements relate \cite{benthic2025mei,structjumper,Schanzer:2019:AccessibleAST}. Structural reasoning requires understanding how elements in a data structure relate (\eg{} parent–child links, left–right orientation, depth, adjacency) and applying those relationships to verify data structure properties or simulate algorithms.}

\edit{Our Wizard-of-Oz study revealed that BVI learners can only maintain positional awareness---and thus reason accurately about these relationships---when overviews are clearly separated from structural detail (\DR{4}). When data structures are presented as diagrams, they encode computational properties that ALT text alone cannot convey, as screen readers typically only access them as static, non-interactive text. Tasks such as verifying BST properties require simultaneously tracking a node’s value, its left and right children, and its depth---relationships that are difficult to reconstruct without explicit positional cues. In the second study, several participants preferred tactile graphics for tree-based tasks because the tactile layouts made positional relationships easier to interpret, whereas some digital formats made those cues harder to identify.}
\edit{Accordingly, accessible representations should preserve explicit structural information even if doing so increases navigational effort. Explicit context helps users understand how a data structure is organized. In practice, this includes consistent spatial layouts in tactile forms, explicit positional cues in digital representations (e.g., ``left child of,'' ``parent of,'' rather than only ``connected to''), and maintaining a clear separation between structural specification and explanatory commentary.}

\paragraph{DP4. Robust - Consistency Across Modalities.}
\edit{WCAG’s \textit{Robust} principle emphasizes that content should remain consistent and reliable across assistive technologies, rather than imposing a specific access modality (or not supporting any). Most prior work on accessible data structures has largely focused on \textit{translation}---converting visual diagrams directly into tactile graphics through improved spacing, labeling, or layout standards \cite{improving-understanding}. Similar patterns appear in visualization accessibility, where modalities are typically treated as one-to-one conversions (\eg visual-to-audio, visual-to-tactile) \cite{wimer2024viz,kim2021accessible}. These approaches improve access, but they inherit the assumptions and constraints of the visual source because each new modality begins with the diagram’s appearance rather than its underlying structure. }

\edit{Our first study showed that individual prototypes conveyed different subsets of the structure (\DR{5}). In the second study, several participants moved deliberately across representations---checking spatial relationships in tactile form, comparing them to digital navigation, and using each modality to clarify what another left uncertain. The representations did not need to be identical; rather, learners benefited when modalities expressed the same underlying relationships, even if said relationships were presented differently; participants emphasized that the representations ``built on each other,'' which helped them reconcile gaps across formats. }
\edit{Our findings suggest that the central issue is not consistency across modalities by itself, but the starting point for translation. When translation begins from a visual diagram, each modality can diverge slightly because it interprets the visual layout differently. Accessible representations can be more robust when they originate from a structural description (\ie{} a diagram specification or intermediate representation) rather than the visual form. A structure-first model ensures that all modalities draw from the same semantics, reducing opportunities for divergence as new formats are introduced.}
\cut{Preserving semantic clarity, structural reasoning, and enabling navigation are not merely matters of usability—they are determinants of whether students can engage in the same conceptual practices as their peers. When computational meaning remains intact and consistent across accessible representations, BVI learners can participate in algorithmic reasoning, structural verification, and mental model construction at the same level as sighted students.}

\subsection{Designing for Scalable, Instructor-Led Accessibility}
\edit{A persistent gap in accessible CS education is that most prior work addresses learner-facing accessibility---supporting BVI students directly---without equipping educators to author accessible materials themselves \cite{blaser2018equity,snodgrass2016instructional,kanchi2025systems,ladner2015increasing,ladner2016all}. This divide is increasingly consequential as a 2024 Department of Justice rule about Title II of the Americans with Disabilities Act places institutional responsibility on instructors to deliver accessible content at scale \cite{DOJ}, despite limited expertise among instructors \cite{bong2024increasing}. \arbor lowers adoption barriers and treats accessibility as an extension of existing curricular practices. This design choice---meeting educators where they are---suggests a broader opportunity: accessibility systems that leverage disciplinary conventions can make accessible authoring feel seamless.}

\edit{\arbors design promotes consistent data structure diagrams across large instructional teams by providing a unified interface for learners, regardless of the diagram specification instructors use. Because instructors and teaching assistants often create bespoke diagrams with disparate tools, variability in structure and accessibility is common. Standardizing output representations is therefore essential, as existing tools rarely preserve the computational properties of data structures or support sufficient accessible alternatives. By relying on familiar specification languages, \arbor also offers a low barrier to entry: instructors and TAs who know these languages can begin authoring diagrams immediately and adapt the tool to their workflows with minimal overhead. \arbor also supports rapid authoring in ad-hoc settings such as office hours, where instructors can generate examples on the fly---for instance, creating a custom binary tree to illustrate the BST property for a student.}





\edit{Our multimodal output strategy further supports scalability across learner needs and institutional contexts. By generating multiple representations from a single source, \arbor can support novices---who can access the representations in combination to form a strong mental model---and more advanced learners who can efficiently utilize the representation that best suits them once spatial relationships become internalized.}
\edit{\arbors design highlights broader principles: scalable accessibility in CS education emerges not from perfecting single-modality representations, but from systems that (1) minimize the instructional workload required to create accessible content and (2) support learners across expertise levels through synchronized multimodal representations.}

\subsection{Limitations}
While \arbor demonstrates the feasibility of generating multiple accessible representations of data structures, several limitations point to directions for future work. Our evaluation was necessarily introductory, focusing on foundational concepts rather than complex algorithmic tasks. Future studies should examine how these principles extend to advanced data structures such as linked lists, graphs, and multidimensional arrays, as well as to more sophisticated reasoning tasks. In addition, our remote study design limited direct observation of tactile interactions, underscoring the value of in-person studies that can capture the nuanced ways learners engage with physical representations.

\section{Future Work and Conclusions}
\edit{Diagrams are foundational to computing education, but they remain difficult to access nonvisually because accessibility practices are often grounded in visual renderings rather than the underlying structural information in the diagram. Our work reframes this challenge by identifying the loss of computational properties---such as roles, relationships, and ordering--- as a key barrier when diagrams are presented as static images or text descriptions. Through two studies, we examined where nonvisual representations fall short and derived design principles that specify what structure-aligned accessibility should provide. \arbor operationalizes these principles by starting from a diagram specification rather than a visual form. By generating synchronized tabular, navigable, and tactile representations from a shared structural model, it supports educators in automatically generating multimodal representations for their data structure diagrams. Because all formats draw from the same underlying model, updates made once can be reflected across every representation, supporting more consistent materials and making it feasible to create, maintain, and update accessible diagrams at the pace required for classes and instructional workflows.}


\edit{Looking forward, our principles suggest opportunities to extend structure-aligned accessibility beyond data structures. Many visually-intensive domains with complex graphics (\eg{} graphs, process diagrams, scientific visualizations) share the same challenges of implicit semantics, spatial reasoning, and dependence on visual layout. Future work can explore how structural model authoring can support multimodal access across these domains.}

\edit{Ultimately, accessible learning at scale requires tools that integrate accessibility into instructional design itself, rather than layering accommodations on top of visual materials. By grounding each representation in a shared structural model rather than the visual form of the diagram, \arbor demonstrates how accessibility can be made systematic, scalable, and integral to computing education.}

\begin{acks}
We thank Ather Sharif, Sean Mealin, and Richard E. Ladner for formative contributions to this work, and Gaby de Jongh and Scott Ferguson at the University of Washington’s Access Technology Center for their guidance. We also thank the UW CSE 12X instructors, teaching assistants, and students who motivated this research. This work was supported in part by the NSF BPC Alliance AccessComputing (2137312), the Paul G. Allen School of Computer Science \& Engineering, and a Google PhD Fellowship, as well as a grant from the National Institute on Disability, Independent Living, and Rehabilitation Research (NIDILRR grant number 90REGE0026-01-00) funding the Center for Research and Education on Accessible Technology and Experiences (CREATE). NIDILRR is a Center within the Administration for Community Living (ACL), Department of Health and Human Services (HHS). The contents of this work do not necessarily represent the policy of NIDILRR, ACL, HHS, and you should not assume endorsement by the Federal Government.
\end{acks}
\bibliographystyle{ACM-Reference-Format}
\bibliography{base.bib}

\end{CJK*}

\end{document}